\documentclass{jfm}
\usepackage{graphicx}
\usepackage{epstopdf, epsfig}
\usepackage{color}
\usepackage{amssymb}
\usepackage{amsmath}
\usepackage{tabularx}
\usepackage{upgreek}
\usepackage{esint}
\usepackage{todonotes}

\newcommand{\la}{\left\langle}
\newcommand{\ra}{\right\rangle}
\newcommand{\be}{\begin{equation}}
\newcommand{\ee}{\end{equation}}
\newcommand{\bse}{\begin{subequations}}
\newcommand{\ese}{\end{subequations}}
\newcommand{\bea}{\begin{eqnarray}}
\newcommand{\eea}{\end{eqnarray}}
\newcommand{\ba}{\begin{array}}
\newcommand{\ea}{\end{array}}

\shorttitle{Insights into energy transfers in turbulence}
\shortauthor{M. K. Verma}

\title{Insights into the Energy Transfers in Hydrodynamic Turbulence Using Field-theoretic Tools}

\author{Mahendra K. Verma\aff{1}
  \corresp{\email{mkv@iitk.ac.in}}}

\affiliation{\aff{1}Department of Physics, Indian Institute of Technology Kanpur, Kanpur 208016, India}

\begin{document}

\maketitle

\begin{abstract}
Turbulent flows exhibit intriguing energy transfers. In this paper, we compute the renormalized viscosities, mode-to-mode energy transfers, energy fluxes, and shell-to-shell energy transfers for the  two-dimensional (2D) and three-dimensional (3D) hydrodynamic turbulence (HDT) using field-theoretic methods. We employ Craya-Herring basis that provides separate renormalized viscosities and energy transfers for its two components. In addition, Craya-Herring basis eliminates complex tensor algebra and   simplifies the calculations considerably. 

In the $k^{-5/3}$ spectral regime of 2D HDT, the energy transfers between neighbouring (\textit{local})  wavenumbers are forward, but they are backwards for distant (\textit{nonlocal})  wavenumbers. The individual transfers between the distant wavenumber shells are small, but their cumulative sum is significant and it overcomes the forward local transfer to yield a constant inverse energy cascade.

For 3D HDT, the mode-to-mode and shell-to-shell energy transfers  reveal forward energy transfers for both  local and  nonlocal wavenumbers. More importantly, using scale-by-scale energy transfers we show that the cumulative nonlocal and local energy  transfers are of the same order, which is contrary to the assumption of \textit{local energy transfers} in turbulence. 

For 3D HDT, the renormalized viscosity, $\nu_2(k)$, computed by us and other authors are in general agreement, and it varies as $k^{-4/3}$. For 2D HDT, our the renormalized viscosity, $\nu_1(k)>0$, but $\nu_1(k)$ reported in literature shows significant variations including a negative $\nu_1(k)$.  In this paper, we argue that the inconsistencies in $\nu_1(k)$ indicate inadequacy of renormalization group analysis that takes into account only local interactions and excludes  nonlocal ones. In 2D HDT, the opposing nature of local and nonlocal energy transfers amplifies the error in $\nu_1(k)$. Based on the \textit{sweeping effect}, which is connected to the  nonlocal interactions, \cite{Kraichnan:PF1964Eulerian} too advocated nonapplicability of field theory in the Eulerian framework.

\end{abstract}


\section{Introduction}

Field theory provides powerful tools that have been used to explain complex phenomena in high-energy physics, condensed-matter physics, statistical physics, and turbulence~\citep{Peskin:book:QFT,Lancaster:book}. This paper focusses on the  application of field theory to the \textit{locality of interactions} and \textit{viscosity renormalization} in  two-dimensional (2D) and three-dimensional (3D) hydrodynamic turbulence (HDT).  Throughout the paper, we assume the flow to be incompressible.

Thousands of papers have been written on the field-theoretic analysis of turbulence. These works provide valuable insights into the dynamics of HDT, magnetohydrodynamic (MHD) turbulence, scalar turbulence, weak turbulence, rotating turbulence, etc. In particular, we learn about the  renormalized parameters, energy fluxes, and correlation functions using field theory. Yet, there are many unresolved issues, e.g., intermittency, higher-order correlations, nature of multiscale energy transfers, application of field theory to anisotropic turbulence, etc.  Therefore,  researchers  still use field theory to solve the above problems. Note that numerical simulations and experiments complement such analytical works.

Kolmogorov, Taylor, Batchelor, Chandrasekhar, Onsager, and others pioneered analytical theories of turbulence, some of which may also classify as field theory. However, to limit the scope of the paper, we skip the works of the above giants and focus on the extension of quantum field theory to turbulence, which was pioneered by Kraichnan. In his first work, \citet{Kraichnan:JFM1959} employed first-order expansion, which he termed as \textit{Direct Interaction Approximation} (DIA), to 3D HDT and derived expressions for the Green's function and correlation function. Kraichnan employed a wavenumber cutoff for an integral for the Green's function that led to a prediction of $k^{-3/2}$ energy spectrum, which  is not observed in experiments.  Later, \cite{Kraichnan:PF1964Eulerian} discovered \textit{sweeping effect}, which is a transport of small-scale fluctuations by the large-scale structures; in this work, Kraichnan argued that  field theory cannot be applied to HDT in the Eulerian framework because of the sweeping effect.

These negative results prompted  Kraichnan to employ Lagrangian-based field theory to turbulence that predicts $k^{-5/3}$ energy spectrum~\citep{Kraichnan:PF1964Lagrangian_Eulerian}. However, in the Eulerian framework itself,  \cite{Kraichnan:JFM1971_2D3D} computed the \textit{effective viscosity} of the Green's function and the  Kolmogorov constants for 2D and 3D HDT. Here, Kraichnan reported that the effective viscosities for 2D HDT and 3D HDT are positive.  However, based on energy fluxes, \cite{Kraichnan:JAS1976} reported \textit{negative eddy viscosity} for 2D HDT, which is opposite to his earlier prediction~\citep{Kraichnan:JFM1971_2D3D}. These contradictions are due to the complexities of 2D HDT, which will be discussed in this paper.  Refer to \cite{Leslie:book} for a detailed discussion on Kraichnan's works on turbulence.

The effective viscosity of the Green's function is computed rigorously using renormalization group (RG) analysis that does not require a wavenumber cutoff. The leading RG calculations for 3D HDT are by \citet{Forster:PRA1977}, \citet{Yakhot:JSC1986}, \citet{DeDominicis:PRA1979}, \citet{Orszag:CP1973}, \citet{McComb:PRA1983},  \citet{McComb:book:HIT}, \citet{Zhou:PRA1988}, and \citet{Zhou:PR2010}  who predicted the renormalized viscosity to scale as $k^{-4/3}$. For the RG calculation, many authors  employed wavenumber-dependent forcing  and noise renormalization \citep{Forster:PRA1977,Yakhot:JSC1986}.  On the other hand, McComb, Zhou, and coworkers used self-consistent recursive RG scheme for their computation.  The other notable field-theoretic works on turbulence are by \cite{Wyld:AP1961}, \cite{Martin:PRA1973}, \cite{Eyink:PRE1993},	\cite{Moriconi:PRE2004}, \cite{Canet:JFM2022}, \cite{Arad:PRE1999}, \cite{Biferale:PR2005}, \cite{Bos:JFM2013},  \cite{Kaneda:FDR2007}, among others.

The works on the RG analysis of 2D HDT is limited (in comparison to 3D HDT). In one of the leading works,  \citet{Olla:PRL1991}  obtained two different spectral regimes by tuning the noise: $k^{-3}$ energy spectrum with a  constant enstrophy flux, and $k^{-5/3}$ spectrum with a constant   energy flux.  For the $k^{-5/3}$ spectral regime, \citet{Olla:PRL1991} derived the renormalization constant to be 0.642 and the Kolmogorov constant to be 6.45. \citet{Nandy:IJMPB1995}  employed self-consistent mode-coupling scheme and obtained similar constants.
\citet{Liang:PF1993} argued that no RG  fixed point exists for 2D HDT due to the dual cascade.

There are significant number of works on the field-theoretic computation of energy flux. \cite{Kraichnan:JFM1959} was the first researcher to formulate  field-theoretic computation of energy flux using \textit{combined energy transfer} and DIA.
 Tools such as Quasi Normal Approximation,  Eddy-damped Quasi Normal Markovian Approximation (EDQNM), and Lagrangian-based DIA were developed by \cite{Kraichnan:JFM1959}, \cite{Chandrasekhar:PRSA1955}, \cite{Herring:PF1965}, \cite{Orszag:CP1973}, \cite{Leslie:book}, and others. Other notable field-theoretic computations of energy flux are by \cite{Yakhot:JSC1986}, \cite{McComb:book:Turbulence}, and \cite{Zhou:PRA1989}.  Refer to \cite{Leslie:book}, \cite{Lesieur:book:Turbulence}, and \cite{McComb:book:Turbulence} for more details.

Later, a new formalism called \textit{mode-to-mode energy transfer} was developed to compute the energy fluxes, as well as shell-to-shell energy transfers~\citep{Dar:PD2001}.  This formalism clearly identifies the \textit{giver} and \textit{receiver} Fourier modes   in a triad, thus resolves the ambiguity encountered in the computation of  shell-to-shell energy transfers using the combined energy transfer~\citep{Domaradzki:PF1990,	Dar:PD2001,Verma:PR2004}.  \cite{Verma:PRE2001} and \cite{Verma:PR2004} employed field theory to compute the energy fluxes for HDT and MHD turbulence using  the mode-to-mode energy transfers.

Starting from~\cite{Kolmogorov:DANS1941Structure}, it has been assumed that the energy transfers in HDT is \textit{local} in Fourier space. That is, the most dominant energy transfers occur among the wavenumbers of similar magnitudes. \cite{Domaradzki:PF1990}, \cite{Verma:Pramana2005S2S}, \cite{Domaradzki:PF2009}, and \cite{Aluie:PF2009} examined this hypothesis by studying the shell-to-shell energy transfers using EDQNM approximation, numerical simulations, and field-theoretic tools. They observed significant nonlocal energy transfers in 2D  and 3D HDT. However, they reported that the cumulative local transfers dominate the nonlocal counterpart because the  \textit{local triads} outnumber the nonlocal ones.  In this paper, we show that the above result is incorrect, and that the cumulative local and nonlocal energy transfers are of the same order.

The field-theoretic tools are very powerful, but they are too complex and daunting involving complicated tensor algebra and singular integrals. Fortunately, we discovered that Craya-Herring basis ~\citep{Craya:thesis,Herring:PF1974,Sagaut:book,Verma:book:ET} simplifies the field-theoretic calculations of turbulent flows  dramatically.  Another major benefit of this basis is that it allows a separate computation of the renormalized viscosities and energy transfers for each component (one component in 2D, and two components in 3D).  As a result, using Craya-Herring basis we not only obtain finer details of turbulence, but also eliminate the complex tensor algebra. We will demonstrate these simplifications in this paper.  In addition, we deviate from the conventional $\int dpdq \delta({\bf k-p-q})$ integrals to $\int dp d\gamma$, where $\gamma$ is the angle between \textbf{k} and \textbf{p} in a triad $(k,p,q)$. This new scheme simplifies the asymptotic analysis, as well as the evaluation of the singular integrals of energy fluxes.

Now, we brief the  results of this paper in relation to the past works.  In this paper, we  employ perturbative field theory to first order and compute the renormalized viscosities, mode-to-mode energy transfers, energy fluxes, and shell-to-shell energy transfers for 2D and 3D HDT. We  perform our calculation in Craya-Herring basis that provides separate contributions from the two components of Craya-Herring basis.  For the RG analysis, we employ recursive procedure proposed by McComb, Zhou, and coworkers~\citep{McComb:PRA1983,Zhou:PRA1989,McComb:book:Turbulence,Zhou:PR2010}.  For the energy transfer computation, we employ Quasi-normal approximation, as in \cite{Kraichnan:JFM1959} and \cite{Orszag:CP1973}. 

Using the above method, we  compute the energy fluxes and the Kolmogorov constants for 2D and 3D HDT in the wavenumber regime where the energy spectra are $k^{-5/3}$. Our calculation, which is in Craya-Herring basis, is much more concise than the previous works.    We show that the energy flux is positive for 3D HDT, but reversed for 2D HDT. These results are consistent with earlier works.

Using field theory we compute the mode-to-mode energy transfer rates for 2D and 3D triads.  For a triad $(k,p,q)$ with $k$ as the receiver wavenumber and $p$ as giver wavenumber, we show that in 2D HDT, the energy transfer is forward when $p \lessapprox k $ (local wavenumbers), but it is backward for distant wavenumbers, i.e., when $p \ll k$. In 3D HDT, the energy transfers are  forward for both local   and distant wavenumbers. The shell-to-shell energy transfers  exhibit similar behaviour. Further, the shell-to-shell energy transfer between distant shells decreases as the shell separation increases; this variation follows $(K/P)^{-4/3}$ scaling where $K$ and $P$ are the wavenumbers of the receiver and giver shells respectively.  These results are consistent with those of \citet{Domaradzki:PF1992} and \citet{Verma:Pramana2005S2S}.

In this paper we go a step further and compute the cumulative local and nonlocal  transfers  by summing up the energy transfers over appropriate wavenumber bands.  Using detailed study we show that for 3D HDT, the cumulative local and nonlocal energy transfers are of the same order. For example, the energy transfers   $(0.7 R, R) \rightarrow (R, R/0.7)$ and  $(0, 0.1 R) \rightarrow (10 R, \infty)$ contribute approximately 17\% each to the total energy flux $\Pi(R)$ for a wavenumber sphere of radius $R$; here $(k_1,k_2)$ represents a wavenumber band.  This result is contrary to the general belief that the eenergy transfers in turbulence is local in wavenumber space. Our work negates the conclusions of \citet{Verma:Pramana2005S2S} and \cite{Aluie:PF2009} who had argued that the local energy transfers dominate the nonlocal ones because the  local triads outnumber the nonlocal ones.

In 2D HDT, the energy transfers between neighbouring shells is forward and significant, but the transfer between the distant shells is backward and small. However, a sum of nonlocal energy transfers is large and it overcompensates the forward energy transfer,  yielding an inverse energy flux.   

We also compute the renormalized viscosity, $\nu_1(k)$, for 2D  HDT, and find this to be positive and approximately 1/4th of the reported values by \cite{Kraichnan:JFM1971_2D3D}, \cite{Olla:PRL1991}, and \cite{Nandi:PRL2014}.  Researchers have reported negative  $\nu_1(k)$ or absence of stable RG fixed point for 2D HDT~\citep{Kraichnan:JAS1976,Verma:Pramana2005S2S,Liang:PF1993}. In this paper, we argue that the inconsistencies in  $\nu_1(k)$ are due to the inability of the RG analysis to take into account the  nonlocal interactions. The coarse-graining process in RG considers the local energy transfers only, which is forward; this is the reason why  $\nu_1(k)$ is positive in our calculation.  The eddy viscosity based on the energy flux is negative because the  energy flux  includes both local and nonlocal interactions   \citep{Kraichnan:JAS1976}. 

The velocity field for 3D HDT has two Craya-Herring  components. Our 3D  RG calculation shows that the renormalized viscosity for the transverse component is positive, and it dominates the renormalized viscosity for the parallel component.  In 3D HDT, the RG parameter has no inconsistency (in contrast to 2D HDT) because the local and nonlocal energy transfers are positive for this case.

The outline of the paper is as follows: In Sections 2, we introduce the relevant hydrodynamic equations in Fourier space and in Craya-Herring basis. In Section 3, we describe the renormalization group analysis of 2D  and 3D HDT. Sections 4 and 5 contain discussions on the energy transfers in a triad, as well as the energy flux in 2D and 3D HDT.  The shell-to-shell energy transfers in 2D and 3D HDT have been discussed in Sections 6 and 7 respectively. We conclude in Section 8.

\section{Governing Equations}

In this section, we introduce the hydrodynamic equations in Fourier space, as well as in Craya-Herring basis.  We also define the correlation functions, energy spectra, and volume integrals for HDT. These quantities play an important role in  perturbative field theory of turbulence, which is a theme of this paper.

\subsection{Hydrodynamics in Spectral Space}
The  Navier--Stokes equation describing  an incompressible flow   is
\bea
\frac{\partial {\bf u}({\bf r},t)}{\partial t} + {\bf u}({\bf r},t) \cdot \nabla {\bf u}({\bf r},t)  &= &
- \nabla p({\bf r},t)  + \nu \nabla^2 {\bf u}({\bf r},t) + {\bf F}_u({\bf r},t), \label{eq:NS_R} \\
\nabla \cdot \mathbf{u}({\bf r},t) & = & 0, \label{eq:div_u_eq_0}
\eea
where ${\bf u}$ is the velocity fluctuation with a zero mean, ${\bf F}_u$ is the external force,  $p$ is the pressure field, and $\nu$ is the kinematic viscosity.  Without a loss of generality, we assume  the density of the fluid to be unity, which leads to the constraint of  Eq.~(\ref{eq:div_u_eq_0}).

In this paper, we  solve the Fourier-version of Eqs.~(\ref{eq:NS_R}, \ref{eq:div_u_eq_0}) using field theory.   In field theory, it is often assumed that the system is of infinite extent, where the transformation from real space to Fourier space and vice versa are as follows~\citep{Peskin:book:QFT}:
\bea
{\bf u}({\bf r},t) & = & \int \frac{d\bf k}{(2\pi)^d} {\bf u}({\bf k}, t) \exp( i {\bf k \cdot r}), 
\label{eq:Fourier_ktox} \\
{\bf u}({\bf k}, t) & = &   \int d{\bf r} [{\bf u}({\bf r},t)   \exp( -i {\bf k \cdot r})],
\label{eq:Fourier_xtok}
\eea
where $d$ is the space dimension, which could be 2 or 3 in this paper.  In Fourier space, Navier-Stokes equations are~\citep{Lesieur:book:Turbulence,Verma:book:ET}
\bea
(\partial_t +\nu k^2){\bf u} (\mathbf{k},t) 
& = & - i \int \frac{d\bf k}{(2\pi)^d}    \{ {\bf k} \cdot  {\bf u}({\bf q}, t) \} {\bf u}({\bf p}, t)    -i {\bf k} p (\mathbf{k},t)   + {\bf F}_u({\bf k},t) , \label{eq:uk}\\
{\bf k\cdot u} (\mathbf{k},t) & = & 0, \label{eq:k_uk_zero}
\eea 
where ${\bf k = p+q}$. The pressure field is determined by taking dot product of Eq.~(\ref{eq:uk}) with ${\bf k}$, that is,
\be 
p(\mathbf{k},t) =  - \frac{i}{k^2} {\bf k} \cdot {\bf F}_u({\bf k},t)  - \frac{1}{k^2}  \int \frac{d\bf k}{(2\pi)^d}    \{ {\bf k} \cdot  {\bf u}({\bf q}, t) \}   \{ {\bf k} \cdot  {\bf u}({\bf p}, t) \} \}  .
\label{eq:Pk}
\ee

It is straightforward to derive the following evolution equation for the \textit{modal energy} $E({\bf k}) = |{\bf u(k)}|^2/2$~\citep{Dar:PD2001,Verma:PR2004}:
\bea
(\partial_t +2 \nu k^2) E(\mathbf{k},t) & = &   \sum_{\bf p}  S^{uu}({\bf k|p|q})  + \Re[ {\bf F}_u({\bf k},t) \cdot {\bf u}^*({\bf k}, t) ]  ,
\label{eq:Euk}
\eea
where
\be
S^{uu}({\bf k|p|q}) = \Im \left[   \{ {\bf k} \cdot  {\bf u}({\bf q}, t) \}   \{ {\bf u}({\bf p}, t) \cdot  {\bf u}^*({\bf k}, t) \} \}    \right]
\label{eq:Suu}
\ee
is the \textit{mode-to-mode energy transfer rate} from the \textit{giver} mode ${\bf u(p)}$ to the \textit{receiver} mode ${\bf u(k)}$ with the mediation of mode ${\bf u(q)}$.  Using $S^{uu}({\bf k|p|q})$, we  compute the \textit{energy flux} emanating from a wavenumber sphere of radius $R$ as follows.  The energy flux $ \Pi(R)$ is the net nonlinear energy transfer rate from all the modes residing inside the sphere to the modes outside the sphere. Hence, the  ensemble average of  $ \Pi(R)$ is~\citep{Dar:PD2001,Verma:PR2004,Verma:book:ET}
\be
\la  \Pi(R)  \ra = \int_{R}^\infty \frac{d{\bf k'}}{(2\pi)^d} \int_0^{R} \frac{d{\bf p}}{(2\pi)^d}  \la S^{u u}({\bf k'|p|q}) \ra.
\label{eq:fluid_flux}
\ee

In this paper, we will compute the renormalized viscosity, as well as $ \la S^{u u}({\bf k'|p|q}) \ra$ and  $\la \Pi(k_0) \ra$, using field theory.  In general, these calculations are quite complex, but they get simplified in Craya-Herring basis,   which will be introduced in the next subsection.

\subsection{Hydrodynamics in Craya-Herring Basis}
\label{sec:HD_CHbasis}

\begin{figure}
	\begin{center}
		\includegraphics[scale = 0.65]{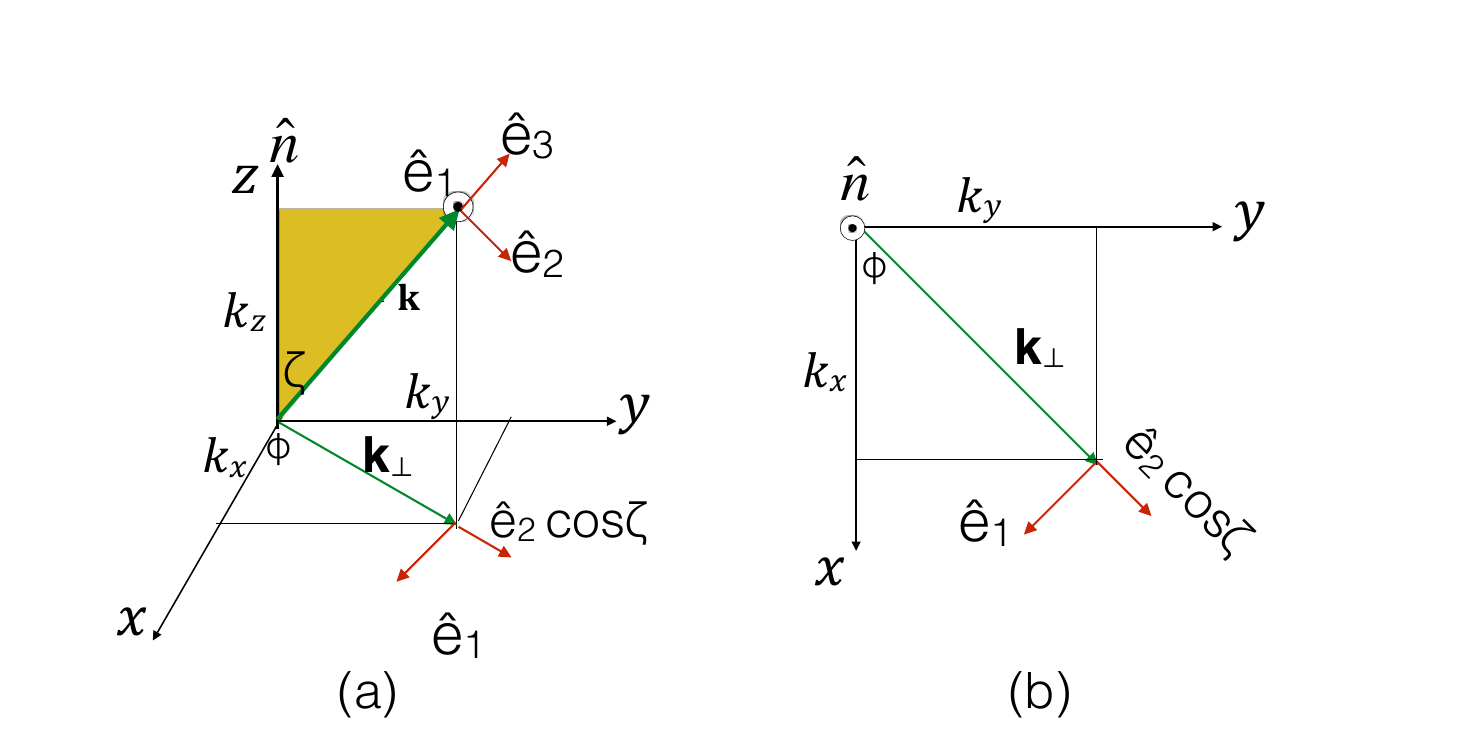}
	\end{center}
	\vspace*{0pt}
	\caption{In 3D: (a) Craya-Herring basis vectors $\{ \hat{e}_1({\bf k}), \hat{e}_3({\bf k}), \hat{e}_3({\bf k}) \}$. (b) A projection of the basis vectors on the $(k_x,k_y)$ plane.   From \cite{Verma:book:ET}. Reprinted with permission from Verma and Cambridge University Press.}
	\label{fig:CH_basis}
\end{figure}

In this section, we briefly describe the relevant hydrodynamic equations in Craya-Herring basis. The basis vectors, illustrated in Fig.~\ref{fig:CH_basis}, are~\citep{Craya:thesis,Herring:PF1974,Sagaut:book}:
\bea
\hat{e}_3({\bf k}) = \hat{k};~~
\hat{e}_1({\bf k}) = \frac{ \hat{k} \times \hat{n}}{|\hat{k} \times \hat{n}|};~~~
\hat{e}_2({\bf k})= \hat{e}_3({\bf k})  \times \hat{e}_1({\bf k}) ,   \label{eq:CH_basis_defn}
\eea 
where  the unit vector $\hat{k}$ is along the wavenumber ${\bf k}$, and the unit vector $\hat{n}$ is chosen along any direction. For incompressible flows, 
\bea
{\bf u}({\bf k},t) & = & u_1({\bf k},t)  \hat{e}_1 + u_2({\bf k},t)   \hat{e}_2.%\label{eq:uk_CH_decomp}
\eea

\begin{figure}
	\begin{center}
		\includegraphics[scale = 1]{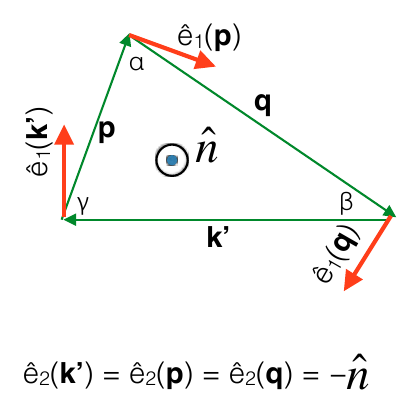}
	\end{center}
	\vspace*{0pt}
	\caption{(a) Craya-Herring basis vectors  for an interacting wavenumber triad. From \cite{Verma:book:ET}, Reprinted with permission from Verma and Cambridge University Press.}
	\label{fig:CH_triad}
\end{figure}

In this paper, we will derive the renormalized viscosity and energy flux by summing up contributions from all the interacting triads. Therefore, as a first step, we write down the evolution equations for  $u_1({\bf k},t)$ and $u_2({\bf k},t)$ in a triad.  For the same, we consider a wavenumber triad $({\bf k',p,q})$  with $({\bf k'+p+q}) = 0$, and choose $\hat{n} $ as follows~\citep{Waleffe:PF1992,Verma:book:ET}:
\be
\hat{n} = \frac{{\bf q \times p}} {|{\bf q \times p}|}.
\label{eq:hat_n}
\ee
Since ${\bf k = p+q}$, we deduce that ${\bf k' = -k}$.  The Craya-Herring basis vectors for the interacting wavenumbers   are illustrated in Fig.~\ref{fig:CH_triad}.  Note that $ \alpha, \beta$,  $\gamma$ are the angles in front of $ k, p, q $ respectively. 

\citet{Verma:book:ET} derived the equations for $u_1$ and $u_2$ components of the triad shown in Fig.~\ref{fig:CH_triad} as
\bea
(\partial_t +\nu k^2){u}_1({\bf k'},t) & = & i k' \sin(\beta-\gamma) u_1^*({\bf p},t)  u_1^*({\bf q},t) +F_1({\bf k'},t) ,
\label{eq:u1k_dot} \\
(\partial_t +\nu k^2){u}_1({\bf p},t) & = &i p \sin(\gamma-\alpha) u_1^*({\bf q},t) u_1^*({\bf k'},t)  +F_1({\bf p},t)  , 
\label{eq:u1p_dot}\\
(\partial_t +\nu k^2){u}_1({\bf q},t) & = &i q \sin(\alpha - \beta)   u_1^*({\bf p},t) u_1^*({\bf k'},t) +F_1({\bf q},t) , 
\label{eq:u1q_dot}
\eea 
\bea
(\partial_t +\nu k^2){u}_2({\bf k'},t) & = &i k'  \{ \sin \gamma u_1^*({\bf p},t)  u_2^*({\bf q},t) -\sin\beta u_1^*({\bf q},t)  u_2^*({\bf p},t)\} +F_2({\bf k'},t) , \label{eq:u2k_dot} \\
(\partial_t +\nu k^2){u}_2({\bf p},t) & = &i p  \{ \sin \alpha u_1^*({\bf q},t)  u_2^*({\bf k'},t) -\sin\gamma u_1^*({\bf k'},t)  u_2^*({\bf q},t)\} + F_2({\bf p},t) , \label{eq:u2p_dot} \\
(\partial_t +\nu k^2){u}_2({\bf q},t) & = &i q  \{ \sin \beta u_1^*({\bf k'},t)  u_2^*({\bf p},t) -\sin\alpha u_1^*({\bf p},t)  u_2^*({\bf k'},t)\} +F_2({\bf q},t) .
\label{eq:u2q_dot}
\eea 
In this paper we show that the field-theoretic treatment of the above equations is much more compact in comparison with their cartesian counterparts.  

In terms of Craya-Herring vectors, the modal kinetic energy $E({\bf k})$ and modal kinetic helicity  $H_K ({\bf k})$ in 3D are
\bea
E ({\bf k},t) & = & \frac{1}{2} \left( |u_1({\bf k},t)|^2   + |u_2({\bf k},t)|^2 \right) =  C_1({\bf k},t) + C_2({\bf k},t).  \label{eq:CH_Ek} \\
H_K ({\bf k},t) & = & \frac{1}{2} \Re[ {\bf u}^*({\bf k},t) \cdot \boldsymbol{\omega}({\bf k},t)  ]  = k \Im[u_1^*({\bf k},t)  u_2({\bf k},t)].  \label{eq:CH_HK}
\eea
where $\boldsymbol{\omega}({\bf k},t) = i k \left[  - u_2({\bf k},t) \hat{e}_1({\bf k}) + u_1({\bf k},t) \hat{e}_2({\bf k})  \right]$ is the vorticity field.  In 2D, $C_2({\bf k},t)=0$ and $H_K ({\bf k},t) =0$.

Nonlinear interactions introduce energy transfers among the Fourier modes. These transfers are compactly captured by the following mode-to-mode energy transfers~\citep{Dar:PD2001,Verma:book:ET}, 
\bea
S^{uu}({\bf k'|p|q}) & = &
S^{u_1 u_1}({\bf k'|p|q}) + S^{u_2 u_2}({\bf k'|p|q}),
\label{eq:Suu_kpq_CH}
\eea
with
\bea
S^{u_1 u_1}({\bf k'|p|q})& = &k' \sin\beta \cos \gamma  \Im \{u_1({\bf q},t) u_1({\bf p},t) u_1({\bf k'},t) \} ,  \label{eq:Su1u1} \\
S^{u_2 u_2}({\bf k'|p|q}) & = &- k' \sin\beta \Im \{u_1({\bf q},t) u_2({\bf p},t) u_2({\bf k'},t) \} .\label{eq:Su2u2}
\eea
In Craya-Herring basis, the energy transfers are either along the $u_1$ channel or along the $u_2$ channel.  These observations not only provide valuable insights into the energy transfers, but they also make the calculations compact in comparison to cartesian basis.  Also note that the dynamical and energy transfer equations in Craya-Herring basis do not involve pressure and complex tensors. 

\subsection{Energy Spectrum and Volume Integrals}

 In this paper, we will focus on homogeneous and isotropic hydrodynamic turbulence in 2D and 3D. However, it is insightful to construct correlation functions for a $d$-dimensional flow:
\be
\la u_i({\bf k}, t) u_j({\bf k'}, t)  \ra = \left[P_{ij}({\bf k}) C_u({\bf k}) - i \epsilon_{ijk} k_k \frac{H_K({\bf k})}{k^2} \right] \delta({\bf k+k'}),
\label{eq:uk_uk'_corr}
\ee
where $C_u({\bf k}) $ is the equal-time velocity correlation function, $H_K({\bf k})$ is the kinetic helicity, and $ \epsilon_{ijk}$ is the Levi-Civita tensor.  In this paper, we focus on nonhelical flows where $H_K({\bf k}) = 0$.  An isotropic $d$-dimensional divergence-free vector field has $d-1$ Craya-Herring  components with  
\be
\la |u_1({\bf k}|^2 \ra = \la |u_2({\bf k}|^2 \ra = ... = \la |u_{d-1}({\bf k}|^2 \ra   = C_u({\bf k}).
\ee
The total kinetic energy is
\bea 
\frac{\la u^2 \ra}{2} = \int E(k) dk = \frac{1}{2} \int  \frac{d{\bf k }} {(2\pi)^{d}} (d-1) C_u({\bf k}) = \frac{1}{2} \frac{S_d}{(2\pi)^d} (d-1) \int dk  k^{d-1} C_u({\bf k}), 
\eea
where $E(k)$ is the one-dimensional (1D) shell spectrum, and $S_d = 2 \pi^{d/2}/\Gamma(d/2)$ is the surface area of the $d$-dimensional sphere.  Specifically, $S_3 = 4\pi$ and $S_2 = 2 \pi$. Using Eq.~(\ref{eq:uk_uk'_corr}), we derive the following  relationship between the modal energy and 1D energy spectrum~\citep{Kraichnan:JFM1959,Leslie:book,Verma:PR2004}:
\be
E(k) = \frac{(d-1)}{2} C_u({\bf k}) \frac{S_d  k^{d-1} }{ (2\pi)^d}.
\ee
Hence, in 3D, 
\be
C_u({\bf k})  =  (2\pi)^3 \frac{E(k)}{4\pi k^2},
\label{eq:Cu_3D}
\ee
and in 2D,
\be
C_u({\bf k})  =  (2\pi)^2 \frac{E(k)}{\pi k}.
\label{eq:Cu_2D}
\ee

\begin{figure}
	\begin{center}
		\includegraphics[scale = 0.65]{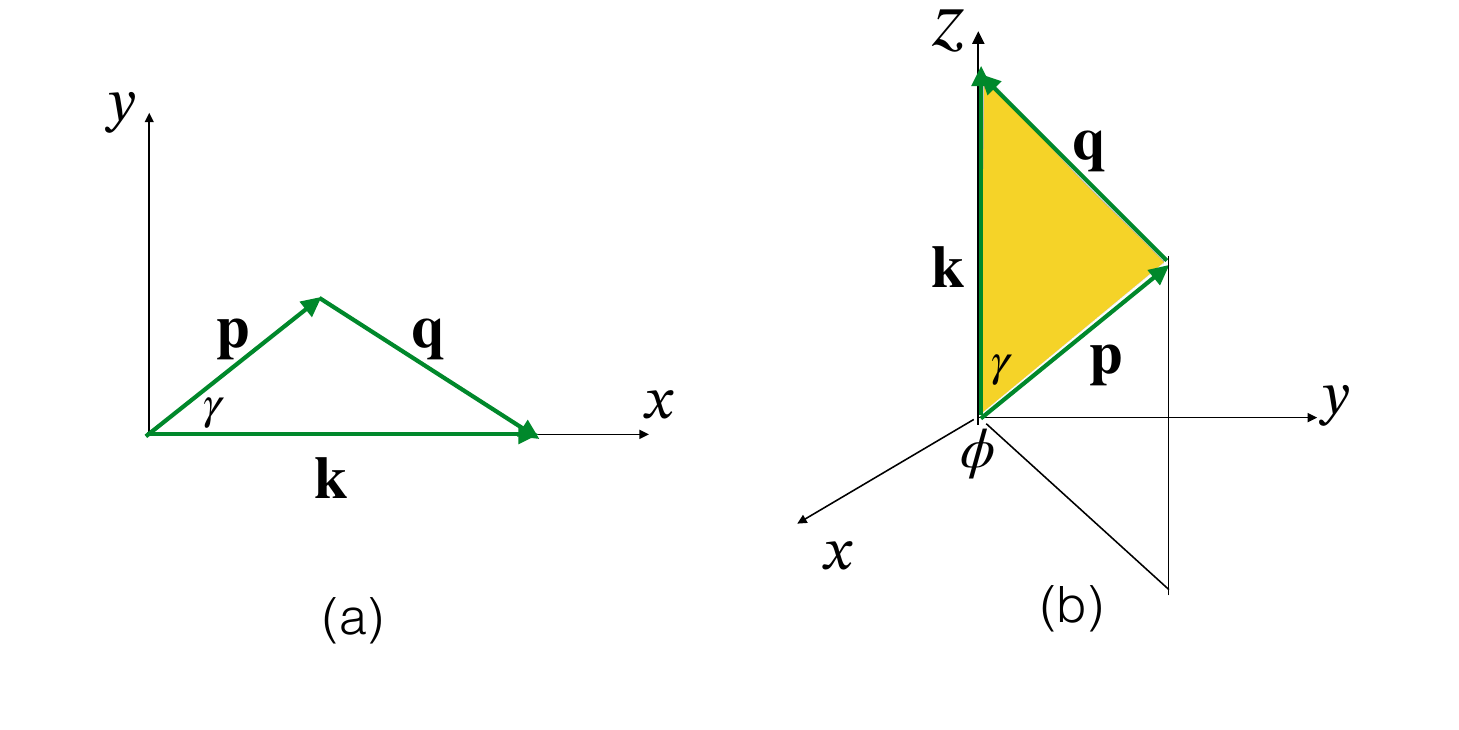}
	\end{center}
	\vspace*{0pt}
	\caption{Illustrations of the volume integration in (a) 2D and (b) 3D.  In 2D, we vary $p$ and $\gamma$ for a given ${\bf k}$.  In 3D, $p$, $\gamma$, and $\phi$ are varied for a given ${\bf k}$. Note that ${\bf q = k-p}$ is fixed for a given ${\bf k}$ and ${\bf p}$. }
	\label{fig:volume}
\end{figure}

Field-theoretic computations of turbulence involve volume integrals, which are typically of the form $ \int d{\bf p} d{\bf q}  \delta({\bf k-p-q})$ for a given $ {\bf k} $ ~\citep{Peskin:book:QFT, Kraichnan:JFM1959,Leslie:book}. Evaluation of such integrals is complex due to the constraint  $ {\bf k = p+q} $~\citep{Leslie:book}.  In this paper, we replace the above integrals with $ \int d{\bf p}  \sim \int p^{d-1} dp d(\cos \gamma)$, where $\gamma$ is the angle between ${\bf k}$ and ${\bf p}$ (see Fig.~\ref{fig:CH_triad} and Fig.~\ref{fig:volume}).  Note that ${\bf q = k-p}$ is fixed for a given $ {\bf k} $ and $ {\bf p} $.  The latter integrals are more intuitive and easier to compute than the former one ($ \int d{\bf p} d{\bf q}  \delta({\bf k-p-q}))$, as will be explained later.

 Refer to Fig.~\ref{fig:volume} for an illustration.    For a 2D integral, we vary $p$ and $\gamma$.  We observe that the integrands for renormalization and energy transfers are symmetric under $\gamma \rightarrow -\gamma$. Hence, we employ
\be
\int_0^{2\pi}  d\gamma \rightarrow 2 \int_0^{\pi}  d\gamma. 
\ee
On the contrary, for the 3D integral, we vary $p$, $\gamma$, and $\phi$.  The azimuthal symmetry yields
\be
\int_0^{2\pi} d\phi = 2 \pi.
\ee

As illustrated in Eq.~(\ref{eq:fluid_flux}), the energy-flux integral involves two sums: (a) $\int d {\bf p}$ representing the energy transfers to ${\bf u(k)}$  from all ${\bf u(p)}$'s residing inside the sphere of radius $R$, and (b)$\int d {\bf k}$  representing a sum over all ${\bf u(k)}$'s present outside the sphere.   Hence, the  integrals for the energy fluxes are of the form
\bea
 \int d\mathrm{VolET} =  \int d{\bf k} \int d{\bf p} =  \begin{cases}
	(2\pi \times 2) \int_{R}^\infty  k dk  \int_0^{R} p dp  \int_{-1}^1 \frac{dz}{\sqrt{1-z^2}} \text{~~~for~$d=2$} \\
	(4\pi \times 2\pi) \int_{R}^\infty  k^2 dk  \int_0^{R} p^2 dp  \int_{-1}^1 dz \text{~~~for~$d=3$}
\end{cases}
\label{eq:dvolET_defn}
\eea  
where $z= \cos \gamma$, and $d\gamma = dz/\sin \gamma$.

For ease of integration of Eq.~(\ref{eq:dvolET_defn}), following \citet{Kraichnan:JFM1959} and \citet{Leslie:book},  we employ  the following transformation for a triad $(k,p,q)$ (see Fig.~\ref{fig:CH_triad}):
\be
k = \frac{R}{u};~~~p = \frac{R v}{u};~~~q = \frac{R w}{u},
\label{eq:uvw_transform}
\ee
for which 
\be
dk dp = \frac{k^2}{u} du dv.
\ee
Using the Jacobian $k^2/u$, we convert the 3D integral to
\be
\int d\mathrm{VolET} = 8 \pi^2 k^6
\int_0^1 \frac{du}{u} \int_0^u dv \int_{-1}^1 dz
=  8 \pi^2 k^6
\int_0^1 dv \int_v^1 \frac{du}{u} \int_{-1}^1 dz.
\ee
Interestingly, the energy-flux integrand for an isotropic flow depends only on $v $ and $z$, and not on $u$. Hence, following \citet{Kraichnan:JFM1959} and  \citet{Leslie:book}, we derive that
\be
\int d\mathrm{VolET} =   8 \pi^2 k^6
\int_0^1 dv  [ \log(1/v)] v^2 \int_{-1}^1 dz.
\label{eq:dVolET_3D}
\ee
For a 2D flow, the corresponding integral is
\be
\int d\mathrm{VolET} =    4\pi  k^4
\int_0^1 dv   [\log(1/v)] v \int_{-1}^1  \frac{dz}{\sqrt{1-z^2}}.
\label{eq:dVolET_2D}
\ee

In our integration scheme, the $ dz  $ integral has fixed limits, $-1$ and 1, in contrast to variable limits in  $ \int_{|k-p|}^{k+p} dq $, which is a part of  $ \int d{\bf p} d{\bf q}  \delta({\bf k-p-q})$.  The fixed limits for the integrals provide certain advantages. As we show in Sections~\ref{sec:Pik_2D} and \ref{sec:Pik_3D},  the $\int_{-1}^1 dz f_1(v,z)$  is conveniently solved using the Gauss-Jacobi quadrature. An evaluation of the corresponding integral $ \int_{|k-p|}^{k+p} dq f_2(p,q) $ using Gaussian quadrature is more complex. In addition, the fixed limits provide simplification in the asymptotic analysis, as well as  in plotting.   Thus, our integration scheme  is more intuitive and simpler than the earlier one ($ \int d{\bf p} d{\bf q}  \delta({\bf k-p-q})$).

The integrals for the RG operations have the following form. The  coarse-graining  operation during renormalization constrains the $p$ and $z$ variables. Hence, in 2D, 
\be
\int d\mathrm{VolRG} = \int d{\bf p} 
=  2 \int_{k_n}^{k_n b}  p dp  \int_{\mathrm{lower}}^\mathrm{upper} \frac{dz}{\sqrt{1-z^2}}, 
\label{eq:dVolRG_2D}
\ee
whereas in 3D, 
\be
\int d\mathrm{VolRG} =\int d{\bf p} =   
2\pi \int_{k_n}^{k_n b} p^2 dp  \int_{\mathrm{lower}}^\mathrm{upper} dz,
\label{eq:dVolRG_3D}
\ee  
where $p \in (k_n, k_n b)$. We will provide more details on the  integration limits in Section~\ref{sec:RG}. 

After the above preliminary discussion on the relevant equations and correlation function, etc., we are ready to perform field-theoretic calculation, namely renormalization group  and energy transfer analysis.  There are many works on the renormalization group analysis of turbulence, hence, we will cover this topic briefly. However, the energy transfer calculations are detailed with several important and novel results.

\section{Renormalization Group Analysis of Hydrodynamic Turbulence}
\label{sec:RG}

In this section, we derive the renormalized viscosities for 2D HDT and 3D HDT using Craya-Herring basis that simplifies the calculations considerably.   We follow the recursive RG method proposed by McComb, Zhou, and coworkers~\citep{McComb:PRA1983,McComb:book:Turbulence,Zhou:PRA1988}.  Note that the coupling constant, the coefficient in front of the nonlinear term ${\bf u \cdot \nabla u}$, is unchanged under renormalization due to the Galilean invariance~\citep{Forster:PRA1977,McComb:book:HIT}.  Therefore,  vertex renormalization is not required in HDT. In addition, in the recursive RG, the forcing or noise is introduced at large scales so as to produce a steady-state with Kolmogorov spectrum.  Hence \textit{noise renormalization} too is avoided in this scheme, and the energy spectrum is taken as $k^{-5/3}$. Note that such a choice for $E(k)$ is as  arbitrary as the choice of noise  that yields Kolmogorov's spectrum (as is done in noise renormalization).  For further details, we refer the reader to books and reviews---\cite{Leslie:book}, \cite{Lesieur:book:Turbulence},  \cite{McComb:book:Turbulence,McComb:book:HIT}, \cite{Zhou:PR2010,Zhou:PR2021},  \cite{Sagaut:book}, and references therein.

First, we compute the renormalized viscosity for 2D HDT.

\begin{figure}
	\begin{center}
		\includegraphics[scale = 1]{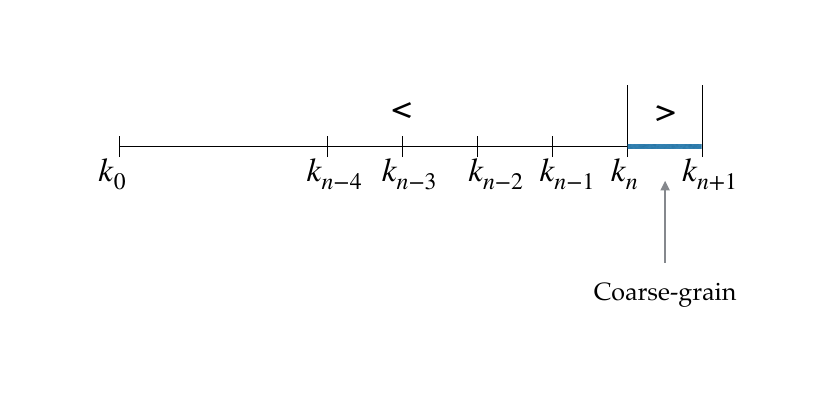}
	\end{center}
	\vspace*{0pt}
	\caption{ In wavenumber renormalization, the modes in the wavenumber band $(k_n, k_{n+1})$, denoted by $>$, are coarse-grained. The coarse-graining leads to enhancement of effective viscosity for wavenumbers $k < k_n$, denoted by $<$. }
	\label{fig:spectrum_flux}
\end{figure}

\subsection{Renormalization of the $u_1$ Component in 2D HDT}
\label{sec:RG_u1}

In 2D HDT, the energy, $ \int d{\bf r}   |{\bf u}|^2$ , and the enstrophy, $ \int d{\bf r} |\nabla \times {\bf u}|^2 $, play an important role~\citep{Lesieur:book:Turbulence}. For a  2D HDT  forced at $k=k_f$, \cite{Kraichnan:PF1967_2D} predicted a constant enstrophy flux for $k>k_f$, and constant energy flux for $k<k_f$ (see  \cite{Boffetta:ARFM2012} and references therein).  In this paper we focus on the latter regime that exhibits $k^{-5/3}$ energy spectrum, and  compute the renormalized viscosity here.

We adopt a recursive renormalization procedure in wavenumber space~\citep{Wilson:PR1974}.  In this scheme, we divide the Fourier space into wavenumber shells $(k_m, k_{m+1})$, where $k_m = k_0 b^m$ with $b$ as a parameter which is greater than unity. We perform \textit{coarse-graining} or averaging operation for a wavenumbers band, and compute its effects on the modes with lower wavenumber. Let us assume that we are at a stage with wavenumbers $(k_0,k_{n+1})$, and the wavenumbers to be coarsegrained are in $(k_n, k_{n+1})$. See Fig.~\ref{fig:spectrum_flux} for an illustration.

We start with a dynamical equation for ${u}_1^< ({\bf k'},t)$ of Eq.~(\ref{eq:u1k_dot}), except that we sum over all ${\bf p}$'s in the convolution [see Eq.~(\ref{eq:uk})]. Note that ${\bf q = -k' -p}$.  In addition, we assume that $F_1({\bf k'},t)$ is active only at large wavenumbers (beyond $k_{n+1}$) that leads to an inverse cascade of energy~\citep{Kraichnan:JFM1971_2D3D}. We assume the flow to be steady due to Ekman-like friction   at small wavenumbers~\citep{Boffetta:ARFM2012}. 

In the dynamical equation for ${u}_1^< ({\bf k'},t)$, the convolution  involves 4 kinds of sums, as illustrated in the following equation:
\bea
[\partial_t + \nu^{(n+1)}_1 k^2 ]{u}_1^< ({\bf k'},t) & = & i k'  \int \frac{d{\bf p}} {(2\pi)^{d}}  \sin(\beta-\gamma) [u_1^{*<}({\bf  p}, t)  u_1^{*<}({\bf q}, t) + u_1^{*<}({\bf  p}, t)  u_1^{*>}({\bf q}, t)] \nonumber \\
&& +  u_1^{*>}({\bf  p}, t)  u_1^{*<}({\bf q}, t)] + u_1^{*>}({\bf  p}, t)  u_1^{*>}({\bf q}, t)] 
\label{eq:u1k_RG0}
\eea
because  $p $ and $q$ may be either less than $k_n$ or greater than $k_n$.  As in Large Eddy Simulations (LES),   $\nu^{(n+1)}_1$ in Eq.~(\ref{eq:u1k_RG0}) represents the renormalized viscosity for $u_1$ when the system wavenumber is $(k_0, k_{n+1})$~\citep{Lesieur:book:Turbulence,Lele:JCP2003}.  Now, we ensemble-average or coarse-grain the fluctuations at scales $(k_n, k_{n+1})$. After this stage of coarse-graining, the viscosity would be $\nu^{(n)}_1$, which acts on the wavenumbers $(k_0,k_{n})$.

For the coarse-graining process,  we assume that  $u_1^{>}({\bf  k}, t) $ is  time-stationary, homogeneous, isotropic, and Gaussian with zero mean, and that $u^<({\bf  k}, t)$  are unaffected by coarse-graining ~\citep{Wilson:PR1974,McComb:book:Turbulence,Zhou:PR2010}.   That is,
\bea
\la u_1^{>}({\bf  k}, t) \ra & = &0, \\
\la u_1^{<}({\bf  k}, t) \ra & = & u_1^{<}({\bf  k}, t).
\eea
Therefore, assuming weak correlation between $<$ and $>$ modes, we arrive at
\bea
\la u_1^{*<}({\bf  p}, t)  u_1^{*<}({\bf q}, t) \ra & = & u_1^{*<}({\bf  p}, t)  u_1^{*<}({\bf q}, t), \\
\la u_1^{*<}({\bf  p}, t)  u_1^{*>}({\bf q}, t) \ra & = &
u_1^{*<}({\bf  p}, t)  \la u_1^{*>}({\bf q}, t) \ra = 0, \\
\la u_1^{*>}({\bf  p}, t)  u_1^{*<}({\bf q}, t) \ra & = &
\la u_1^{*>}({\bf  p}, t)  \ra u_1^{*<}({\bf q}, t) = 0 .
\eea
Substitution of the above relations in Eq.~(\ref{eq:u1k_RG0})  yields
\bea
[\partial_t + \nu^{(n+1)}_1  k^2 ]{u}^<_1({\bf k'},t) & = & i k'  \int \frac{d{\bf p}} {(2\pi)^{d}}  \sin(\beta-\gamma) u_1^{*<}({\bf  p}, t)  u_1^{*<}({\bf q}, t)   \nonumber \\
&&  + i k'  \int_\Delta \frac{d{\bf p}} {(2\pi)^{d}}  \sin(\beta-\gamma)  \la u_1^{*>}({\bf  p}, t)  u_1^{*>}({\bf q}, t) \ra ,
\label{eq:u1k_RG1} 
\eea
where $\Delta$ represents the wavenumber region  $({\bf p, q}) \in (k_{n}, k_{n+1})$.  The second term of  Eq.~(\ref{eq:u1k_RG1}) enhances or renormalizes the kinematic viscosity leading to the following equation:
\bea
[\partial_t + \nu^{(n)}_1  k^2 ]{u}^<_1({\bf k'},t) & = & i k'  \int \frac{d{\bf p}} {(2\pi)^{d}}  \sin(\beta-\gamma) [u_1^{*<}({\bf  p}, t)  u_1^{*<}({\bf q}, t) ],
\label{eq:u1k_RG_final} 
\eea
where 
\be
 \nu^{(n)}_1 k^2 =   \nu^{(n+1)}_1 k^2 - \mathrm{Second~Integral~of~Eq.~(\ref{eq:u1k_RG1})}.
\label{eq:nu_k_second_integral}
\ee

\begin{figure}
	\begin{center}
		\includegraphics[scale = 0.7]{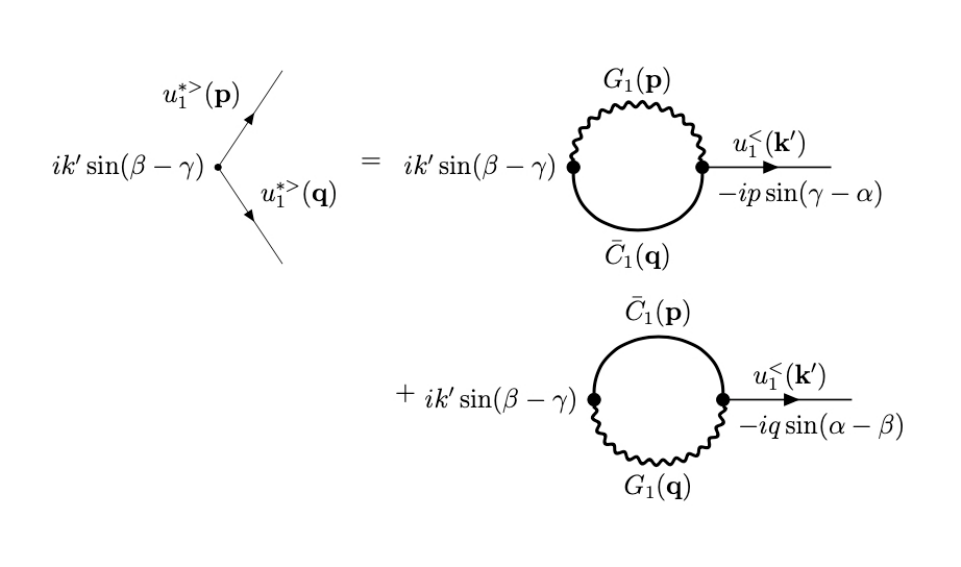}
	\end{center}
	\vspace*{0pt}
	\caption{Feynman diagrams associated with the renormalization of $\nu_1$ for the $u_1$ component.}
	\label{fig:RG_CH1}
\end{figure}

Note that the second integral of Eq.~(\ref{eq:u1k_RG1})  vanishes to the zeroth order. Hence, we expand this terms to first-order in perturbation that leads to   the Feynman diagrams of Fig.~\ref{fig:RG_CH1}.  We compute the integral corresponding to the first Feynman diagram as follows.   We expand $u_1^{*>}({\bf  p}, t)  $ using the Green's function [see Eq.~(\ref{eq:u1p_dot})]:
\bea
u_1^{*>}({\bf  p}, t) = \int_0^t dt' G_1({\bf p},t-t') (-ip) \int \frac{d{\bf r}} {(2\pi)^{d}} \sin(\gamma-\alpha) u_1({\bf r},t') u_1({\bf s},t') ,
\label{eq:u1_p(t)}
\eea
where ${\bf p+ r + s }= 0$.  We substitute the expression of Eq.~(\ref{eq:u1_p(t)}) in the right-hand-side of Eq.~(\ref{eq:u1k_RG1}) and simplify the expression using the following relations~\citep{McComb:book:Turbulence,Zhou:PR2010}:
\bea
\la u_1^*({\bf q}, t) u_1({\bf r}, t) \ra & = & \bar{C}_1({\bf q}, t-t')  \delta({\bf q-r}),
\label{eq:C_q_eq_r} \\
G({\bf k}, t-t') & = & \theta(t-t') \exp[-\nu(k) k^2 (t-t') ],  \label{eq:Gk_tt'} \\
\bar{C}_1({\bf k}, t-t') & = & C_1({\bf k}) \exp[-\nu(k) k^2 (t-t')].  \label{eq:Ck_tt'}
\eea
In the above equations, $\bar{C}_1({\bf k}, t-t') $ is an  unequal time correlation, while $C_1({\bf k}) $ is an equal-time correlation.  Note that $\nu(k)$ of Eqs.~(\ref{eq:Gk_tt'}, \ref{eq:Ck_tt'}) is the renormalized viscosity at wavenumber $k$.  We could compute $\nu(k)$ by coarse-graining modes with wavenumbers beyond $k$.  In this paper, we denote correlation functions corresponding to $u_1$ and $u_2$ as $C_1$ and $C_2$, respectively, in order to contrast the two. Note, however, that $C_1 = C_2$ because of isotropy of the flow. Similar notation is adopted for the Green's functions. 

Interestingly, using Eq.~(\ref{eq:C_q_eq_r}), we obtain
\be 
{\bf s = -p -r = -p -q = k'}.
\ee
Based on the above relations, we deduce the  integral corresponding to the first Feynman diagram as
\bea
I_1 & = &   \int_\Delta \frac{d{\bf p}} {(2\pi)^{d}}  \int_0^t dt' G({\bf p},t-t') (k'p) \sin(\beta-\gamma) \sin(\gamma-\alpha)    C_1({\bf q}, t-t') u^<_1({\bf k'},t').  \nonumber \\
\label{eq:I1_0} 
\eea
Now, we employ Markovian approximation~\citep{Orszag:CP1973,Leslie:book,Lesieur:book:Turbulence}. When $\nu(k) k^2 \gg 1$, the function $\exp[-\nu(k) k^2 (t-t')]$  rises sharply to unity near $t'=t$. Hence, the $dt'$ integral gets maximal contribution near $t'=t$. Therefore, $u_1({\bf k},t') \rightarrow u_1({\bf k},t)$, and
\bea
I_1 & = &   \int_\Delta \frac{d{\bf p}} {(2\pi)^{d}} \frac{kp \sin(\beta-\gamma) \sin(\gamma-\alpha) C_1({\bf q})}{\nu(p)p^2 + \nu(q) q^2} u^<_1({\bf k'},t).
\label{eq:I1} 
\eea

Following similar steps, we compute the integral corresponding to the second Feynman diagram of Fig.~\ref{fig:RG_CH1} as
\bea
I_2 & = &    \int_\Delta \frac{d{\bf p}} {(2\pi)^{d}} \frac{kq \sin(\beta-\gamma) \sin(\alpha-\beta) C_1({\bf p})}{\nu(p)p^2 + \nu(q) q^2} u^<_1({\bf k'},t).
\label{eq:I2} 
\eea
Since $I_1$ and $I_2$ are proportional to $u^<_1({\bf k'},t)$, these terms can added to $\nu^{(n+1)}_1 k^2 u^<_1({\bf k'},t)$ to yield the renormalized viscosity $\nu^{(n)}_1$. In particular, using Eqs.~(\ref{eq:u1k_RG_final}, \ref{eq:nu_k_second_integral}) we show that
\bea
\nu^{(n)}_1 k^2 & = & \nu^{(n+1)}_1 k^2 -  \int_\Delta \frac{d{\bf p}} {(2\pi)^{d}} \frac{k \sin(\beta-\gamma)  }{\nu(p)p^2 + \nu(q) q^2} [p C_1({\bf q}) \sin(\gamma-\alpha) +  q C_1({\bf p}) \sin(\alpha-\beta) ].
\nonumber \\
\label{eq:nu1(k)}
\eea

To compute $\nu^{(n)}_1$,  we choose $k=k_n$ in Eq.~(\ref{eq:nu1(k)}). In addition,   we make the following change of variables:
\be
k = k_n;~~~~~~{\bf p = p'} k_n;~~~~~~{\bf q = q'} k_n
\ee
with $1 \le p' \le b$ and $1 \le q' \le b$ with $b=1.7$ that yields a triad $(1,p',q')$.  We sum over all possible triads to compute the integral of Eq.~(\ref{eq:nu1(k)}).  Using 
\be
q'^2 = 1+p'^2 -2 p' \cos \gamma = 1+p'^2 -2 p'z,
\label{eq:q_p'_z}
\ee
we derive the lower and upper limits of $ z $ for a given $p'$  as
\bea
\mathrm{lower} & = & \frac{p'^2 + 1 -b^2}{2 p'}, \\
\mathrm{upper} & = &  \frac{p'}{2}.
\eea
 We substitute these limits in Eq.~(\ref{eq:nu1(k)}) and compute the integral.  In this paper, we focus on the inverse cascade regime:
\bea
C_1({\bf k}) & = &  (2\pi)^2 \frac{1}{\pi k} E(k),
\label{eq:C1_2d} \\
E(k) & = & K_\mathrm{Ko2D} \epsilon_u^{2/3} k^{-5/3},  \label{eq:Ek_2d} \\
\nu^{(n)}_1 & = & \nu_{1*} \sqrt{K_\mathrm{Ko2D}} \epsilon_u^{1/3} k_n^{-4/3},  \label{eq:nuk_2d}
\eea
where $\epsilon_u$ is the energy flux, $K_\mathrm{Ko2D}$ is the Kolmogorov constant for 2D HDT, and $\nu_{1*}$ is the renormalization constant for 2D HDT in the spectral regime where the  energy flux is constant. Renormalization for the spectral regime with constant enstrophy flux will be presented  in  future.  

We substitute Eqs.~(\ref{eq:C1_2d}-\ref{eq:nuk_2d}) in   Eq.~(\ref{eq:nu1(k)}), and simplify the expressions using trignometric identities for the  triad $(1,p',q')$ (see Fig.~\ref{fig:CH_triad}), e.g.,
\bea
\sin \alpha & = &  \frac{1}{q'} \sin \gamma, \\
\cos \alpha  & = & \frac{p'^2 + q'^2 -1}{2 p' q'} = \frac{p'-z}{q'}.
\eea
At $k=k_n$, these operations yield
\bea
\nu_{1*}   & = & \nu_{1*} b^{-4/3}  - \frac{2}{\pi} \frac{1}{\nu_{1*}} \int_1^b p' dp'  \int^{p'/2}_{(p'^2+1-b^2)/(2p')}  \frac{dz}{\sqrt{1-z^2}} (F_1 + F_2),
\eea
or
\bea
\nu^2_{1*} (1-b^{-4/3})  & = &  - \frac{2}{\pi} \int_1^b p' dp'  \int^{p'/2}_{(p'^2+1-b^2)/(2p')}  \frac{dz}{\sqrt{1-z^2}} (F_1 + F_2),
\label{eq:nu1_integral}
\eea
where the prefactor $2/\pi$ is the results of prefactors of Eqs.~(\ref{eq:Cu_2D},  \ref{eq:dVolRG_2D}), and
\bea
F_1(p',z) & = & \frac{(1-z^2) p' (p'-2z)(2p'z-1) q'^{-14/3}}{p'^{2/3} +q'^{2/3}}, \\
F_2(p',z)  & = & \frac{(1-z^2)  (1-p'^2)(2p'z-1) p'^{-8/3} q'^{-2}}{p'^{2/3} +q'^{2/3}}
\eea
are functions of the independent variables $p'$ and $z$. Note that $q'$ can be expressed in terms of $p'$ and $z$ using Eq.~(\ref{eq:q_p'_z}).  Equation~(\ref{eq:nu1_integral}) differs from those of \cite{McComb:PRA1983} and \cite{Zhou:PRA1989} who computed the correction to $\nu_1(k)$ [Eq.~(\ref{eq:nu1(k)}] for all $k$'s that leads to a $k$-dependent $\nu_{1*}$.  In our paper, we interpret $\nu_1^{(n)}$ as the renormalized viscosity for wavenumbers $(k_0, k_n)$ that leads to a constant $\nu_{1*}$.  Our scheme, which is motivated by LES~\citep{Lesieur:book:Turbulence,Lele:JCP2003}, 
simplifies the computation of $\nu_{1*}$ significantly.  

We compute the integral of Eq.~(\ref{eq:nu1_integral}) numerically.  For    an accurate  integration, we perform the $dz$ integral  using Gaussian quandrature and the  $dp'$ integral  using a Romberg scheme.  Refer to Appendix~\ref{sec:integration} for  details on the integration schemes used in this paper.  We employ Python's {\tt scipy.integrate.romberg} function   whose tolerance limit is $ 1.48 \times 10^{-8} $. Our numerical computation yields  $\nu_{1*}^2 (1-b^{-4/3}) = 0.010527 $, hence,
\be
\nu_{1*} = \sqrt{\frac{0.010527}{1-b^{-4/3}}}= 0.1441.
\label{eq:nu1_0.1026}
\ee
Throughout the paper, we report our results up to 4 significant digits, which is within the tolerance limit. However, we remark that $\nu_{1*} $ depends marginally on $b$~\citep{Zhou:NASA1997}, a topic  which is not  detailed in this paper. 

Let us compare our result with the past ones.   \cite{Kraichnan:JFM1971_2D3D} employed \textit{test field model} to compute $\nu_{1*}$ and observed this to be 0.648. \cite{Olla:PRL1991}'s RG and \cite{Nandy:IJMPB1995}'s mode-coupling theory  predict that $\nu_{1*} \approx 0.612$, similar to that of  \cite{Kraichnan:JFM1971_2D3D}.  In comparison, our prediction of Eq.~(\ref{eq:nu1_0.1026}) is around 4 times smaller.  In addition, because of  the  inverse cascade of energy,  some authors  argue that the \textit{eddy viscosity}, which is same as the renormalized viscosity,  of 2D HDT is negative~\citep{McComb:PRA1983,Zhou:PRA1989} (also see Appendix~\ref{sec:eddy_visc}).  For example, \cite{Kraichnan:JAS1976} reported negative eddy viscosity based on the inverse energy cascade.  \cite{Verma:PR2004} performed recursive RG and predicted that $\nu_{1*} \approx -0.60$, but the convergence to this $\nu_{1*}$ was unsatisfactory.  On the other hand, \cite{Liang:PF1993} reported an absence of RG fixed point due to the dual cascade in 2D HDT.

For 2D HDT, the large variations in the value of $\nu_{1*}$ is confusing, yet they indicate certain limitations of the RG procedure in handling the  complex dynamics of 2D HDT.  The renormalized viscosity $\nu_{1}^{(n)}$ is derived by coarse-graining the  modes in the wavenumber band $(k_n, k_{n+1})$, and by ignoring the modes with longer wavenumbers. Hence,   the RG analysis implicitly assumes locality of interactions.  In contrast, the energy flux calculations include both local and nonlocal interactions. The energy transfers between the local Fourier modes are forward, but those between the nonlocal modes are backward; the latter  transfers are responsible for the inverse energy cascade   [see \cite{Domaradzki:PF1990,Verma:Pramana2005S2S}, and Section~\ref{sec:ETu1}].    This is the reason why the \textit{effective} viscosity based on the  RG analysis is positive, but the eddy viscosity based on energetics is negative~\citep{Kraichnan:JFM1971_2D3D, Olla:PRL1991, Nandy:IJMPB1995, Kraichnan:JAS1976}.  We believe that exclusion of the nonlocal interactions is the reason for the inaccurate RG predictions of $\nu_{1*}$.  Refer to Appendix~\ref{sec:eddy_visc} for details.
 
 %$ which may be the reason for positive $\nu_{1*}(k)$ observed by us,   \cite{Kraichnan:JFM1971_2D3D}, \cite{Olla:PRL1991},  and \cite{Nandy:IJMPB1995}.  However, 
 
%\cite{Kraichnan:JAS1976} analysed the energy equation and derived the eddy viscosity based on the energy flux at wavenumber $k$.  He argued that   the eddy viscosity for 2D HDT is negative due to the inverse  cascade of energy.  Note that the energy flux calculations include both local and nonlocal interactions, which is not the case for the RG analysis.  

 \cite{Kraichnan:PF1964Eulerian}  showed that in HDT, the large-scale structures \textit{sweep} small-scale fluctuations, a phenomena called \textit{sweeping effect}. Based on these observations, Kraichnan argued that Eulerian framework is not suitable for field-theoretic treatment, and advocated for Lagrangian direct interaction approximation (DIA). Since the sweeping effect is related to the nonlocal interactions, our arguments for the inapplicability of RG to 2D turbulence has similarities with those of  \cite{Kraichnan:PF1964Eulerian}.

 The resolution of the above issues  requires detailed theoretical analysis and  numerical simulations.    \cite{Verma:INAE2020_sweeping} and \cite{Verma:PRE2023_shell}  performed direct numerical simulation of 3D HDT and the shell model respectively, and computed $\nu(k)$ using the  unequal-time correlation functions of Eq.~(\ref{eq:Ck_tt'}).  A similar analysis for 2D HDT may provide us the renormalized viscosity that will help validate  the analytical predictions. At the end of this section, we remark that for ease of notation, the renormalized viscosity $\nu^{(n)}_1$ is conveniently written as $\nu_1(k)$; that is,  $\nu_1(k)$ is the effective viscosity for all the wavenumbers less than $k$.

  In the next section, we will compute the renormalized viscosity for 3D HDT.

\subsection{Renormalization of the $u_2$ Component of 3D HDT}
\label{sec:RG_u2}

In 3D HDT, a velocity Fourier mode has two Craya-Herring components, $u_1$ and $u_2$ (see Fig.~\ref{fig:CH_basis}).  In this subsection, we compute the renormalized viscosities for  the two components.  This is unlike past works, e.g., \cite{Yakhot:JSC1986}, \cite{McComb:PRA1983}, and \cite{Zhou:PRA1989}, where a single renormalized viscosity has been computed.  Hence,  Craya-Herring basis provides more details on turbulence dynamics. 

At first, we compute $\nu_2^{(n)}$,  for which we follow the same steps as in Section~\ref{sec:RG_u1}. Note, however, that for 3D HDT, we assume that the flow is forced at large scales to achieve a constant energy flux in the inertial range, where we employ our RG analysis.  We start with Eq.~(\ref{eq:u2k_dot}), but with all possible triads. One of the intermediate steps in the derivation of  $\nu_2^{(n)}$ is
\bea
(\partial_t + \nu_2^{(n+1)} k^2){u}^<_2({\bf k'},t) & = &i k' \int \frac{d{\bf p}} {(2\pi)^{d}}   \{ \sin \gamma u_1^{<*}({\bf p},t)  u_2^{<*}({\bf q},t) -\sin\beta u_1^{<*}({\bf q},t)  u_2^{<*}({\bf p},t)\} \nonumber \\
& & + i k' \int \frac{d{\bf p}} {(2\pi)^{d}}   \{ \sin \gamma u_1^{>*}({\bf p},t)  u_2^{>*}({\bf q},t) -\sin\beta u_1^{>*}({\bf q},t)  u_2^{>*}({\bf p},t)\} ,
\label{eq:u2k_RG} \nonumber \\
\eea
where $\nu_2^{(n+1)}$ is the renormalized viscosity for  $u_2(k)$  with $k \in s(k_0, k_{n+1})$. In the above equation, the terms of the form $\int d{\bf p} \la u_1^{>*}({\bf q},t)  u_2^{>*}({\bf p},t) \ra$ contribute to viscosity renormalization.  In this subsection we show that $\nu_2^{(n+1)} \ne \nu_1^{(n+1)}$, which is expected because the dynamical evolution of $u_1$ and $u_2$  are different. We capture this interesting aspect of viscosity renormalization for the first time.

\begin{figure}
	\begin{center}
		\includegraphics[scale = 0.6]{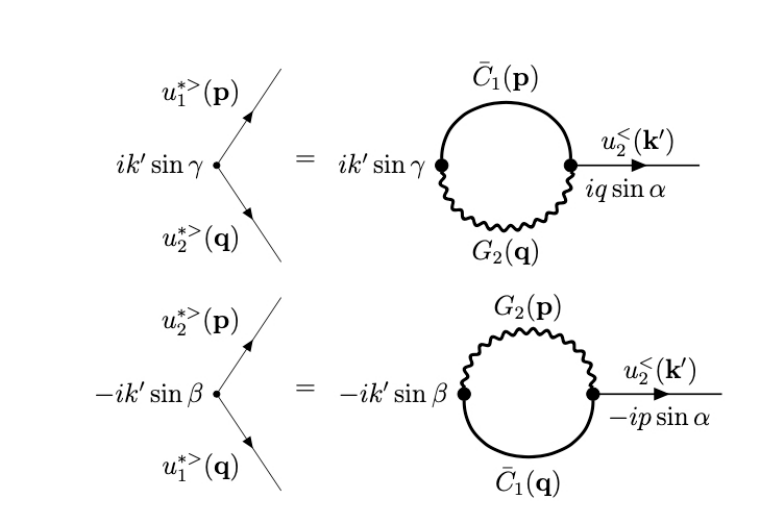}
	\end{center}
	\vspace*{0pt}
	\caption{Feynman diagrams associated with the renormalization of $\nu_2$ for the $u_2$ component. }
	\label{fig:RG_CH2}
\end{figure}

For 3D isotropic HDT, we employ the following scaling laws:
\bea
C_1({\bf k}) & = &  C_2({\bf k})  = (2\pi)^3 \frac{1}{4 \pi k^2} E(k),
\\
E(k) & = & K_\mathrm{Ko} \epsilon_u^{2/3} k^{-5/3}, \\
\nu^{(n)}_2 & = & \nu_{2*} \sqrt{K_\mathrm{Ko}} \epsilon_u^{1/3} k_n^{-4/3}, \label{eq:nu2_defn}
\eea
where $\epsilon_u$ is the energy flux, $K_\mathrm{Ko} $ and $\nu_{2*} $ are the Kolmogorov  and  renormalization constants, respectively, for 3D HDT.
Following similar steps as in the $\nu_{1*}$ computation, we observe that  the second integral of Eq.~(\ref{eq:u2k_RG}) contributes to the viscosity renormalization. The associated Feynman diagrams are shown in Fig.~\ref{fig:RG_CH2}, and the corresponding integral is
\bea
I_3 & = &   - \int_\Delta \frac{d{\bf p}} {(2\pi)^{d}} \frac{kq C_1({\bf p}) \sin \gamma \sin \alpha 
+ kp C_1({\bf q}) \sin \beta \sin \alpha 
}{\nu(p)p^2 + \nu(q) q^2} u_2({\bf k'},t), 
\label{eq:I3} 
\eea
In the nondimensional equation, $I_3$ contributes to  viscosity renormalization    as follows:
\bea
\nu_{2*}^2  (1-b^{-4/3}) & = & \frac{1}{2}  \int_1^b p'^2 dp' 
\int^{p'/2}_{(p'^2+1-b^2)/(2p')} dz F_3 ,
\label{eq:nu2_integral}
\eea
where the prefactor $1/2 = 2\pi/(4\pi)$ is due to the cancellation of factors in Eqs.~(\ref{eq:Cu_3D},  \ref{eq:dVolRG_3D}), and
\bea
F_3(p',z) & = &  (1-z^2)  p'^{-11/3} +(1-z^2) p'^2 q'^{-11/3}.
\eea

To evaluate the integral of Eq.~(\ref{eq:nu2_integral}) we employ Gaussian quadrature and a  Romberg scheme, as in the computation of $\nu_1(k)$.  The numerical computation yields  $I = 0.1197$, hence  
\be
\nu_{2*} = \sqrt{\frac{I}{1-b^{-4/3}}} = 0.4859.
\ee
For 3D HDT,  we compute $\nu_{1*}$  using  a modified version of Eq.~(\ref{eq:nu1_integral}), which is
\bea
\nu_{1*}^2 (1-b^{-4/3}) & = &  \frac{1}{2}  \int_1^b p'^2 dp' 
\int^{p'/2}_{(p'^2+1-b^2)/(2p')} dz (F_4 + F_5)
\label{eq:nu1_3D_integral}
\eea
with
\bea
F_4(p',z) & = & \frac{(1-z^2) p' (p'-2z)(2p'z-1) q'^{-17/3}}{p'^{2/3} +q'^{2/3}}, \\
F_5(p',z)  & = & \frac{(1-z^2)  (1-p'^2)(2p'z-1) p'^{-11/3} q'^{-2}}{p'^{2/3} +q'^{2/3}}.
\eea

A numerical integration of Eq.~(\ref{eq:nu1_3D_integral}) yields
\be
\nu_{1*} = 0.09218.
\label{eq:nu1_for_3D}
\ee
	
Since $\nu_{1*} \ll \nu_{2*}$,  we conclude that for 3D HDT, the $u_2$ component contributes much more to the renormalized viscosity ($\nu_*$) than the $u_1$ component. Hence, we compare $\nu_{2*}$ with the past predictions.  \citet{Kraichnan:JFM1971_2D3D} predicted that $\nu_* = 0.315$, whereas  \citet{McComb:book:Turbulence}  and  \citet{Zhou:PRA1988}
reported  $\nu_{*}$  be approximately 0.40. Other researchers too reported similar values for $\nu_{*}$~\citep{Yakhot:JSC1986,Nandy:IJMPB1995}. \citet{Zhou:PRA1988} argued that an  inclusion of triple nonlinearity in the RG analysis increases $\nu_{*}$  to around 0.5, but this topic beyond the scope of this paper. Note that our prediction for $\nu_{2*}$ is comparable to the past ones.

Using Eqs.~(\ref{eq:nuk_2d}, \ref{eq:nu2_defn}), we derive that for both $\nu_1^{(n)}$ and $\nu_2^{(n)}$, 
\be
\frac{\nu_{1,2}^{(n)}}{\nu_{1,2}^{(n)}}
= \left( \frac{k_{n+1}}{k_n} \right)^{-4/3} =  b^{-4/3} .
\ee
In 	quantum field theory,  we express the running coupling constant in terms of $b = \exp(l)$~\citep{Peskin:book:QFT}.  Using $b^{-4/3} \approx 1- 4l/3$ (for small $l$), we derive that
\be
\frac{d \nu}{dl} \approx - \frac{4}{3} \nu.
\ee
Therefore, the renormalized viscosity $\nu$ increases with the decreases of $k_n$. This is similar to the \textit{running coupling constant} in quantum chromodynamics. A cautionary remark, however, is in order.  For  the Navier-Stokes equation, the real coupling constant, which is the coefficient of the nonlinear term, is constant. Here, $\nu$ is similar to the mass term of the $\phi^4$ theory~\citep{Peskin:book:QFT}.

With this we close our discussion on the viscosity renormalization for HDT.  We employed Craya-Herring basis for our computation that eliminates  complex tensor algebra. 	We will employ $\nu_{1*} $ and $\nu_{2*} $  for the computation of energy flux and mode-to-mode energy transfers, which are covered in Sections~\ref{sec:ETu1} and \ref{sec:ETu2}.

\section{Energy Transfers and Flux for the $u_1$ Component and  2D HDT}
\label{sec:ETu1}

Researchers have computed the energy transfers and fluxes for HDT using various field-theoretic tools. Another related topic of interest is \textit{locality of nonlinear interactions} in turbulence.  Note, however, that the energy transfers in 2D HDT has not been studied in detail. This paper attempts to fill this gap by computing  $\la S^{u_1 u_1}({\bf k'|p|q}) \ra$ and the corresponding energy flux using  Craya-Herring basis.  Our calculations are not only compact, but they provide detailed insights into the energy transfers in 2D HDT.

In this section, we compute the energy transfer rates and energy flux in the wavenumber regime with $k^{-5/3}$ energy spectrum. In the Appendix~\ref{sec:Enstrophy}, we briefly discuss the mode-to-mode enstrophy transfer and enstrophy flux in this spectral region.

\subsection{Mode-to-mode Energy Transfers for the $u_1$ Component }
\label{sec:M2M_CH1}

\begin{figure}
	\begin{center}
		\includegraphics[scale = 0.5]{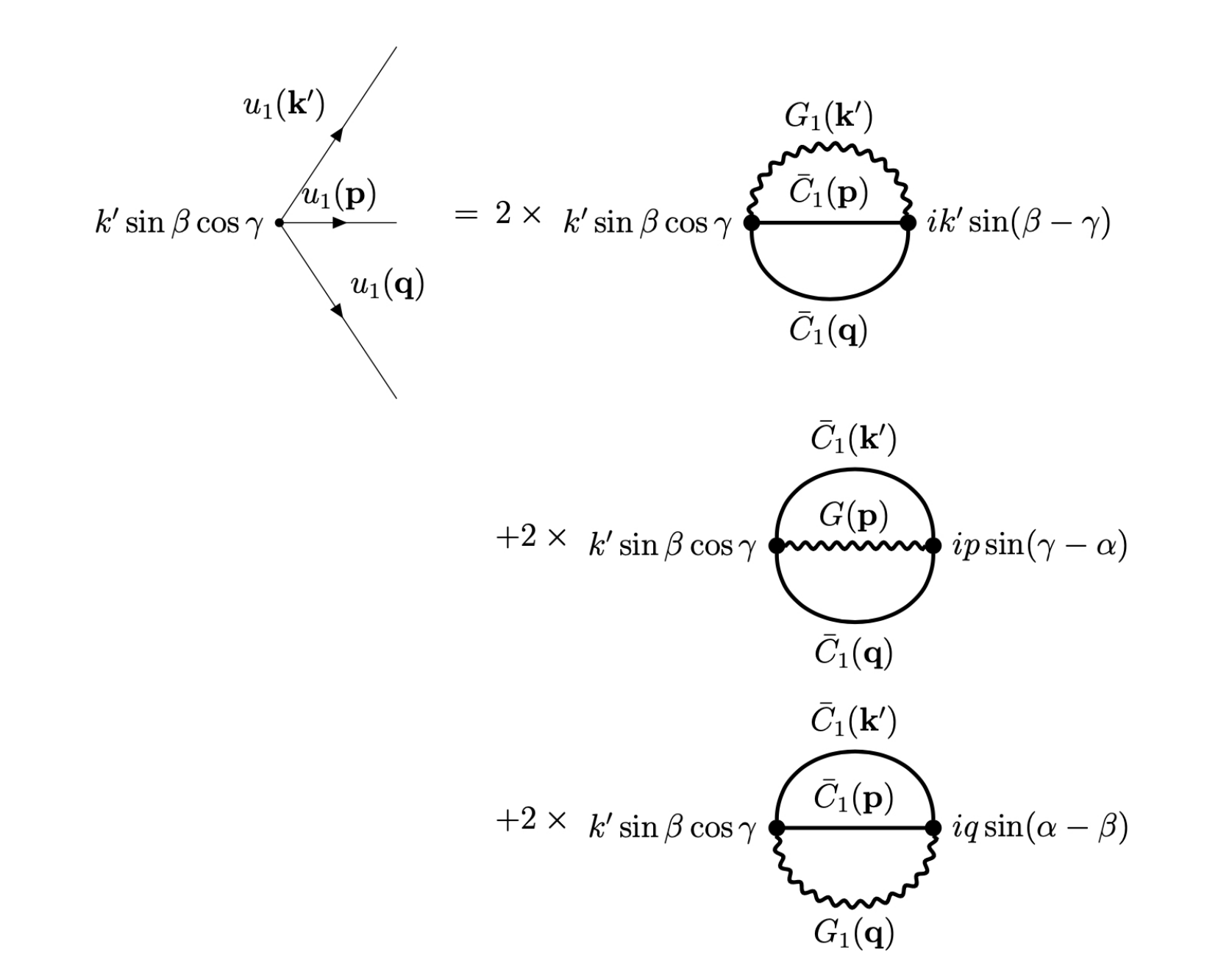}
	\end{center}
	\vspace*{0pt}
	\caption{Feynman diagrams associated with the energy transfers between the $u_1$ components.  Equation~(\ref{eq:four_corr_QN}) illustrates two ways to get the second-order correlation functions, which leads to the factor 2 in all the digrams. }
	\label{fig:ET_CH1}
\end{figure}

In this subsection, we will compute the mode-to-mode energy transfers between  the $u_1$ components of Craya-Herring basis.  We start with Eq.~(\ref{eq:Su1u1}) and present the ensemble-averaged \textit{mode-to-mode energy transfer} from $u_1({\bf p},t)$ to $u_1({\bf k'},t)$  with the mediation of $u_1({\bf q},t)$, which is
\bea
\la S^{u_1 u_1}({\bf k'|p|q}) \ra & = &k' \sin\beta \cos \gamma  \Im \{ \la u_1({\bf q},t) u_1({\bf p},t) u_1({\bf k'},t) \ra \}   
\label{eq:Skpq_u1_avg}
\eea
with ${\bf k' + p+q } =0$.  We assume the turbulence to be homogeneous, isotropic, and steady, and that it satisfies Eqs.~(\ref{eq:C1_2d}-\ref{eq:nuk_2d}). To estimate $\la S^{u_1 u_1}({\bf k'|p|q}) \ra $, it is customary to assume that the variables $u_1({\bf p},t)$, $u_1({\bf k'},t)$, and $u_1({\bf q},t)$ are quasi-normal, according to which  the three-variable correlation of Eq.~(\ref{eq:Skpq_u1_avg}) is nonzero (unlike Gaussian variables). However, the first-order expansion of the triple correlation of Eq.~(\ref{eq:Skpq_u1_avg}) leads to a fourth-order correlation, that is expanded as a sum of  products of two second-order correlations under the assumption that the variables are Gaussian~\citep{Kraichnan:JFM1959,Orszag:CP1973}.  The corresponding Feynman diagrams are given in  Fig.~\ref{fig:ET_CH1}.

Let us evaluate the integral corresponding to the first Feynman diagram of Fig.~\ref{fig:ET_CH1}. Here, $u_1({\bf k'},t) $ is expanded using the Green's function as [see Eq.~(\ref{eq:u1k_dot})]
\bea
u_1({\bf k'},t) & = & i \int_0^t dt' G_1({\bf k'},t-t') 
 k'  \int \frac{d{\bf r}} {(2\pi)^{d}}  \sin(\beta-\gamma) [u_1^*({\bf  r}, t')  u_1^{*}({\bf s}, t') ]
\eea
with ${\bf k' + r+s } =0$. Substitution of the above in Eq. (\ref{eq:Skpq_u1_avg}) leads to a fourth-order correlation, which is expanded as a sum of products of two second-order correlations:
\begin{align}
\la u_1({\bf q},t) u_1({\bf p},t) u_1({\bf r},t') u_1({\bf s},t') \ra = \la u_1({\bf q},t) u_1({\bf p},t) \ra \la  u_1({\bf r},t') u_1({\bf s},t') \ra  \nonumber + \\
  \la u_1({\bf q},t)  u_1({\bf r},t') \ra \la u_1({\bf p},t) u_1({\bf s},t') \ra + \la u_1({\bf q},t) u_1({\bf s},t')  \ra \la u_1({\bf p},t) u_1({\bf r},t') \ra.
  \label{eq:four_corr_QN}
\end{align}
Note that $\la u_1({\bf q},t) u_1({\bf p},t) \ra  = \la  u_1({\bf r},t') u_1({\bf s},t') \ra = 0$  because ${\bf p} \ne -{\bf q}$ and ${\bf r} \ne -{\bf s}$. Using the above correlations and Eq.~(\ref{eq:uk_uk'_corr}), we deduce that
\bea
\la u_1({\bf q},t) u_1({\bf p},t) u_1({\bf k'},t) \ra_a  
& = &  \int_0^t dt' G_1({\bf k'},t-t') 
i k'  \int \frac{d{\bf p}} {(2\pi)^{d}}  \sin(\beta-\gamma) \times \nonumber  \\
&&  2 [\bar{C}_1({\bf  p},t- t')  \bar{C}_1({\bf q}, t-t') ] .
\eea
Using the properties of temporal relations of Eqs.~(\ref{eq:Gk_tt'}, \ref{eq:Ck_tt'}), we deduce that
\bea
\la u_1({\bf q},t) u_1({\bf p},t) u_1({\bf k'},t) \ra_a  
& = &   \frac{i 2  k' \sin(\beta-\gamma) C({\bf p}) C({\bf q})}{\nu(k) k^2 + \nu(p) p^2 + \nu(q) q^2}.
\eea

\begin{figure}
	\begin{center}
		\includegraphics[scale = 1]{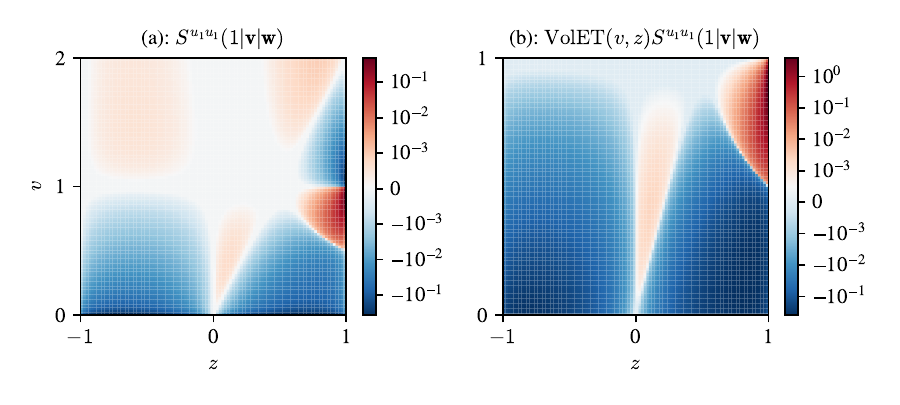}
	\end{center}
	\vspace*{0pt}
	\caption{For energy transfers in 2D HDT: (a) Density plot of $\la S^{u_1 u_1}({\bf k'|p|q}) \ra$. (b)  Density plot of $[v \log(1/v)/\sqrt{1-z^2} ] \la S^{u_1 u_1}({\bf k'|p|q}) \ra$.  }
	\label{fig:Skpq_CH1}
\end{figure}

We compute the integrals corresponding to the other two Feynman diagrams of Fig.~\ref{fig:ET_CH1} and add them to the above, which yields 
\bea
\la S^{u_1 u_1}({\bf k'|p|q}) \ra & = &     \frac{\mathrm{numr}_1 }{\nu(k) k^2 + \nu(p) p^2 + \nu(q) q^2},
\label{eq:Skpq_u1_expanded}
\eea
where
\bea
\mathrm{numr}_1 & = & 2 [k' \sin(\beta-\gamma) C({\bf p})C({\bf q})+p \sin(\gamma-\alpha) C({\bf k'} ) C({\bf q})+q \sin(\alpha-\beta) C({\bf k'} ) C({\bf p}) ]\nonumber \\
&& \times k' \sin\beta \cos \gamma .
\eea
The physics in the inertial range is scale invariant. Hence, we  employ  the transformation of Eq.~(\ref{eq:uvw_transform}) to simplify the integral of Eq.~(\ref{eq:Skpq_u1_expanded}), which leads to
\bea
\la S^{u_1 u_1}({\bf 1|v|w}) \ra(v,z) & = &     \frac{\mathrm{numr}_2 }{ \pi^2 (1+v^{2/3} + w^{2/3})} = k^4 \la S^{u_1 u_1}({\bf k'|p|q}) \ra,
\label{eq:Svw_u1}
\eea
where
\bea
\mathrm{numr}_2 & = &2 w^{-2} [ (2 v^2z^2 - vz) (vw)^{-8/3} + (v^3 z - 2 v^2 z^2)
w^{-8/3} + (1-v^2)  z v^{-5/3} ]
(1-z^2),
\label{eq:numr_Skpq_2D} \nonumber \\
\eea
and $w^2 = 1+ v^2 - 2 v z$.  We study the properties of  mode-to-mode interactions in a triad $(1,v,w)$ using Eqs.~(\ref{eq:Svw_u1}, \ref{eq:numr_Skpq_2D}).  Note that 
$\la S^{u_1 u_1}({\bf k'|p|q}) \ra$ has a  dimension of $k^{-4}$, whereas  $\la S^{u_1 u_1}({\bf 1|v|w}) \ra$ is dimensionless.  In Fig.~\ref{fig:Skpq_CH1}(a) we illustrate the density plot of $\la S^{u_1 u_1}({\bf 1|v|w}) \ra(v,z) $. We deduce the following properties for the energy transfers in 2D HDT using this plot  and  Eqs.~(\ref{eq:Svw_u1}, \ref{eq:numr_Skpq_2D}).

\begin{figure}
	\begin{center}
		\includegraphics[scale = 1]{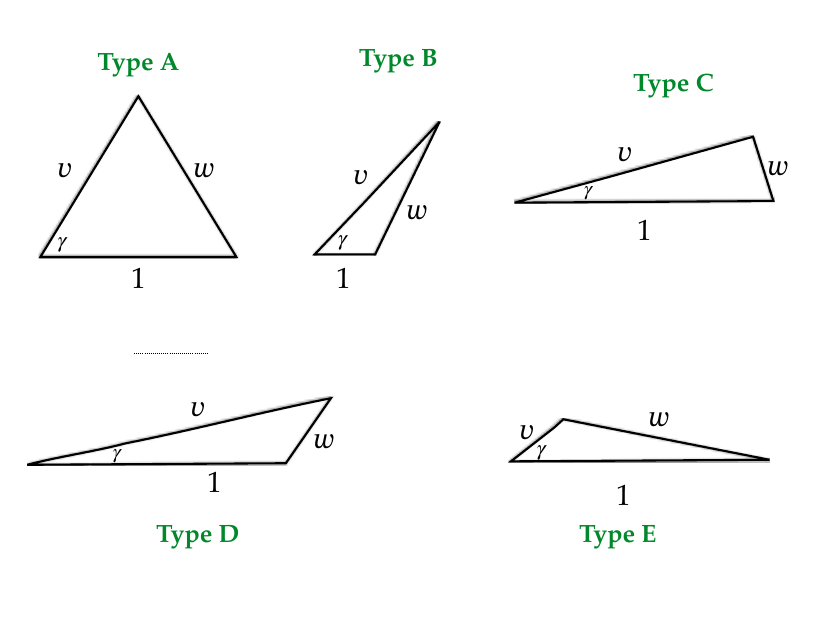}
	\end{center}
	\vspace*{0pt}
	\caption{Various triads involved in energy transfers. The sides of the triangle are $(1,v,w)$, and the angle between the sides 1 and $v$ is $\gamma$. }
	\label{fig:triads}
\end{figure}

It is straightfoward to show that $\la S^{u_1 u_1}({\bf 1|v|w}) \ra = 0$ for an equilateral triad for which $v =w =1$ and $z = 1/2$ (or $\gamma = \pi/3$). This triad is illustrated as Case A in Fig.~\ref{fig:triads}.    Interestingly,  $\la S^{u_1 u_1}({\bf 1|v|w}) \ra \approx 0$ for triads with $v \approx 1$  that corresponds to the central region of Fig.~\ref{fig:Skpq_CH1}(a).   The triads with $v \approx w \gg 1$  too have negligible energy transfers because
\be
\la S^{u_1 u_1}({\bf 1|v|w}) \ra 
\approx -\frac{2}{\pi^2} z^2(1-z^2) v^{-8/3} \rightarrow 0
\ee
in this regime.

In 2D HDT,  the triads with $v \approx 1$ and $z \approx 1$ (Type C and D of Fig.~\ref{fig:triads}), and  the triads with $ v \approx 0$ (Type E)  dominate the energy transfers.  For Type C and D triads, $v \approx 1$, $z \approx 1$, and
$w \approx 0$. Hence, Eq.~(\ref{eq:numr_Skpq_2D}) yields
\bea
\mathrm{numr}_2 & \approx &2  w^{-14/3} [(2v^2 z^2 -vz)v^{-8/3} + v^3 z -2 v^2 z^2] (1-z^2) \nonumber  \\
& \approx &2  w^{-14/3} vz [(2v z (v^{-8/3} -1)  + v^2(1 -v^{-14/3})] (1-z^2) \nonumber  \\
& \approx & 2 w^{-14/3} vz [(2v z (-8/3)(v-1)  + v^2 (14/3) (v-1)] (1-z^2) \nonumber  \\
& \approx & \frac{8}{3} w^{-14/3} (1-v)(1-z).
\eea
Therefore,
\be
\la S^{u_1 u_1}({\bf 1|v|w}) \ra  \approx
\frac{\mathrm{numr}_2 }{2\pi^2} \approx  \frac{4}{3\pi^2}  w^{-14/3} (1-v)(1-z).
\label{eq:Skpq2D_asymptot_typeCD}
\ee
These $\la S^{u_1 u_1}({\bf 1|v|w}) \ra$'s contribute significantly to the energy flux due to $w^{-14/3}$ singularity. Clearly, $\la S^{u_1 u_1}({\bf 1|v|w}) \ra  > 0$ for $v = 1-\epsilon$, and  $\la S^{u_1 u_1}({\bf 1|v|w}) \ra  < 0$ for $v = 1+\epsilon$ ($\epsilon > 0$). Hence, the  energy transfers between the neighbouring wavenumbers is forward, i.e., from smaller wavenumber to larger wavenumber. This feature appears to contradict the inverse energy cascade in the $k^{-5/3}$ spectral regime, but it will be resolved in the following discussion.

Next, we consider the triads of Type E, for which  $v \approx 0$ and $w \approx 1$. For this case, using $w^2  = 1+v^2 -2vz \approx 1 -2vz$, we obtain
\bea
\mathrm{numr}_2 & \approx & 2 v^{-5/3} z [(2vz-1)(1+(8/3) v z) +1-v^2] (1-z^2) \nonumber  \\
& \approx & 2 v^{-5/3} z \left[ -\frac{2}{3} vz +v^2\left( \frac{16}{3} z^2 - 1 \right) \right] (1-z^2) \nonumber  \\
& \approx & -\frac{4}{3} v^{-2/3} z^2 (1-z^2).
\label{eq:2D_v_ll_1_stage0}
\eea
Hence,
\be
\la S^{u_1 u_1}({\bf 1|v|w}) \ra  \approx
\frac{\mathrm{numr}_2 }{2\pi^2} \approx  - \frac{2}{3\pi^2}  v^{-2/3} z^2 (1-z^2)
\label{eq:2D_v_ll_1}
\ee
Therefore, $\la S^{u_1 u_1}({\bf 1|v|w})  \ra  < 0$ for such triads. That is, the modes with small wavenumber  receive energy from $u_1(k=1)$.  These energy transfers lead to an inverse energy transfers in 2D HDT, and they overcompensate the forward local  energy transfers due to the triads with $v \approx 1$.   Interestingly, for triads of Type E, the mode-to-mode enstrophy transfer has mixed sign that leads to a forward enstrophy flux. Refer to Appendix~\ref{sec:Enstrophy} for details.

In Fig.~\ref{fig:Skpq_CH1}(b) we plot  $ \mathrm{VolET} \la S^{u_1 u_1}({\bf 1|v|w}) \ra $, which is   $v \log(1/v)/\sqrt{1-z^2} \la S^{u_1 u_1}({\bf 1|v|w}) \ra $. The two subfigures exhibit similar behaviour, except that the energy transfers in  Fig.~\ref{fig:Skpq_CH1}(b) are   amplified  for the $z \approx 1$ region due to the $1/\sqrt{1-z^2}$ factor.

\subsection{Energy flux for the $u_1$ Component and  2D HDT}
\label{sec:Pik_2D}

In 2D HDT, the energy flux for a wavenumber sphere of radius $R$ is the cumulative energy transfers between the $u_1$ components, as given below:
\be
\la \Pi(R) \ra = \int_{R}^\infty \frac{d{\bf k'}}{(2\pi)^2} \int_0^{R} \frac{d{\bf p}}{(2\pi)^2} \la S^{u_1 u_1}({\bf k'|p|q}) \ra  .
\label{eq:Pi_2d}
\ee
Using  $\la S^{u_1 u_1}({\bf 1|v|w}) \ra $ derived in the previous subsection, we compute the energy flux for 2D HDT in the $k^{-5/3}$ spectral regime.  We substitute $\la S^{u_1 u_1}({\bf 1|v|w}) \ra $  in Eq. (\ref{eq:Pi_2d})), and employ   the change of variables from $k,p,q$ to $u,v,w$ using Eq. (\ref{eq:uvw_transform}).  Consequently,  the energy flux equation gets transformed to
\bea
\frac{ \la \Pi(R)\ra}{\epsilon_u }   =  \frac{ K_\mathrm{Ko2D}^{3/2}}{ \nu_{1*}} I_4, 
\label{eq:Pi_2D_integral0}
\eea
where
\bea
I_4 =  \frac{4}{\pi} 
\int_0^1 dv   [\log(1/v)] v \int_{-1}^1  \frac{dz}{\sqrt{1-z^2}} \frac{\mathrm{numr}_2}{(1+v^{2/3} + w^{2/3})) },
\label{eq:Pi_2D_integral}
\eea
with $\mathrm{numr}_2$ given in Eq.~(\ref{eq:numr_Skpq_2D}), and the prefactor $4/\pi = 4\pi /(\pi^2)$ is due to the cancellation of factors in Eqs.~(\ref{eq:Cu_2D},  \ref{eq:dVolET_3D}).

We compute $I_4$ numerically. To avoid inacurracies due to singularities in $I_4$, we employ the Gauss-Jacobi quadrature for the $dz$ integral, and a Romberg iterative scheme for the $dv$ integral. Refer to Section~\ref{sec:integration} for a brief discussion on the integration procedure.  This process yields $I_4 \approx -0.07569$, whose  negative sign indicates an inverse cascade of energy.  Since $\Pi_u(R)$ of Eq.~(\ref{eq:Pi_2D_integral0}) is independent of $R$, we conclude    that the inertial-range energy flux is constant. Since $\la \Pi(R) \ra = -\epsilon_u $ and $\nu_{1*} = 0.1441$, we estimate the Kolmogorov constant for 2D HDT as
\be
K_\mathrm{Ko2D} =  \left( \frac{\nu_{1*}}{I_4} \right)^{2/3} 
\approx 1.536.
\ee
The above $K_\mathrm{Ko2D}$ differs significantly from the previously reported values of $K_\mathrm{Ko2D}$, which is around  6.   For example, \cite{Kraichnan:JFM1971_2D3D} reported that $K_\mathrm{Ko2D} = 6.69$.  \cite{Olla:PRL1991} and \cite{Nandy:IJMPB1995} reported similar constants in their field-theoretic calculation.  These discrepancies may be due to certain limitations of field-theoretic treatment for 2D HDT, as  discussed in Section~\ref{sec:RG_u1}.

Interestingly, by changing the lower limit of the integral of Eq.~(\ref{eq:Pi_2D_integral}) from $ \int_0^1 dv$ to  $ \int_{0.22}^1 dv$, our formalism yields  $I_4 \approx -0.009773$ and 
\be
\bar{K}_\mathrm{Ko2D} = 6.013,
\ee
which is in a reasonable agreement with the previous results.   The integral $I_4$ is nearly zero for the lower integral limit of 0.265; for this case, $\bar{K}_\mathrm{Ko2D} \rightarrow \infty$. That is,  $\bar{K}_\mathrm{Ko2D}$ increases as the lower limit of $v$  approaches 0.265.  Note that the physical constants depend on integral cutoffs~\citep{Kraichnan:JFM1959,Leslie:book,Peskin:book:QFT}, hence the above adjustment is reasonable.  In the following discussion, we provide a physical interpretation for  the above feature. 

\begin{figure}
	\begin{center}
		\includegraphics[scale = 0.9]{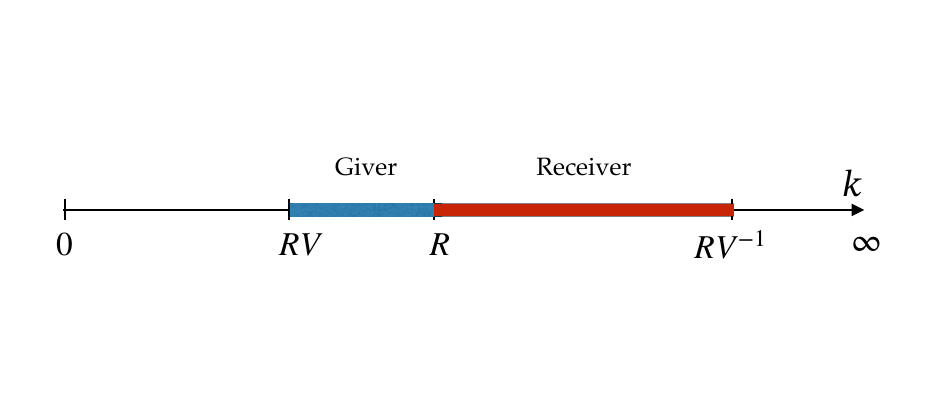}
	\end{center}
	\vspace*{0pt}
	\caption{Consider a wavenumber sphere of radius $R$. The fractional energy flux, $ \Pi_V(R) $, is the net energy transfer from the giver modes in wavenumber band ($R V, R$) to the receiver modes in the band $(R, R V^{-1})$. For $V=0$,  $\Pi_V(R) = \Pi(R) $.}
	\label{fig:Flux_frac}
\end{figure}
\begin{figure}
	\begin{center}
		\includegraphics[scale =0.8]{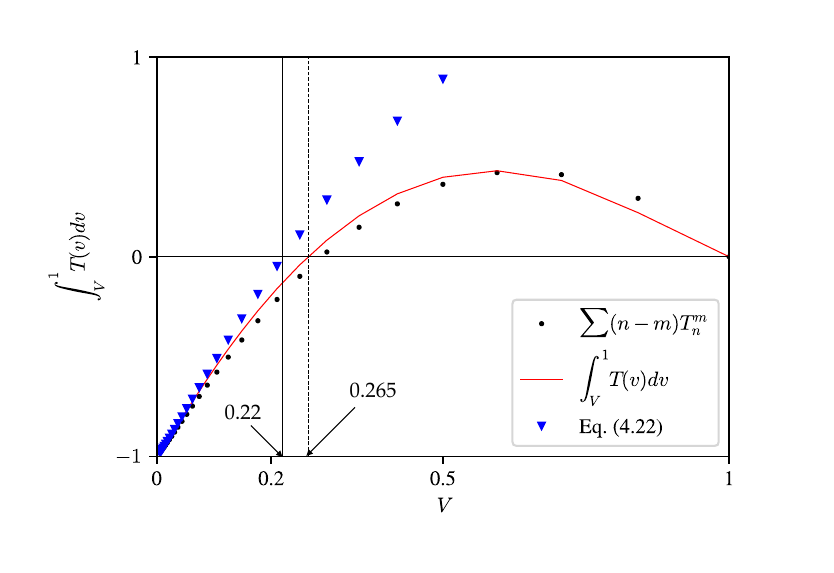}
	\end{center}
	\vspace*{0pt}
\caption{Plots of the fractional energy flux $\int_V^1 T(v) dv$ (red curve), partial sum of shell-to-shell energy transfers (black dots)  [Eq.~(\ref{eq:cumulativeTmn})], and the asymptotic $\int_V^1 T(v) dv$ for small $V$ [Eq.~(4.22)]. The vertical solid line at $V=0.22$  illustrates the lower-limit of integration for which $\bar{K}_\mathrm{Ko2D} = 6.013$. $\int_V^1 T(v) dv \approx 0$ for $V=0.265$, which is at the dashed vertical line.	}
	\label{fig:Flux_fraction_CH1}
\end{figure}

To disentangle the energy transfer contributions from various wavenumber regimes, we define \textit{fractional energy flux},
\bea
\frac{ \la \Pi_V(R)\ra}{\epsilon_u }  &  = & \int_V^1 dv  T(v) \nonumber \\
& = &  \frac{ K_\mathrm{Ko2D}^{3/2}}{\nu_{2*}}  \frac{4}{\pi} 
\int_V^1 dv   [\log(1/v)] v \int_{-1}^1  \frac{dz}{\sqrt{1-z^2}}  
\frac{\mathrm{numr}_2}{(1+v^{2/3} + w^{2/3})  }.
\label{eq:Int_Tv_2D}
\eea
Based on the relations of Eq.~(\ref{eq:uvw_transform}), it is easy to show that $\la \Pi_V(R)\ra$ represents the net energy transfer from the giver modes in the band $(R V,R)$ to the receiver modes in the band $(R, R/V)$. See Fig.~\ref{fig:Flux_frac} for an illustration. 

We choose $V = s^{-n}$ with $s = 2^{1/4}$ and $n$ ranging from 1 to 30. We compute $\int_V^1 dv  T(v)$ for these $V$'s and plot them in Fig.~\ref{fig:Flux_fraction_CH1} that exhibits complex energy transfers in 2D HDT.   First, $\int_{V}^1 dv  T(v) > 0$ for $V \gtrapprox 0.265$. That is, any wavenumber band $(RV, R)$ with $0.265 < V < 1$ transfers energy in the forward direction. Second,  $\int_{0}^1 dv  T(v) = -1$ implying that the modes in the band $(0,0.265 R)$  receive energy via nonlocal interactions and overcompensate the forward energy transfer $\int_{0.265}^1 dv  T(v)$. The latter energy transfers are responsible for the inverse energy cascade.

For small $V$, we can estimate $\int_0^{V} dv  T(v)$ using asymptotic analysis. Since, $v \ll 1$ for this case, we employ Eqs.~(\ref{eq:2D_v_ll_1}, \ref{eq:Int_Tv_2D}) that yields
\bea
\int_0^V dv T(v) & = &  \frac{ K_\mathrm{Ko2D}^{3/2}}{\nu_{1*}}  \frac{4}{\pi} 
\int_V^1 dv [\log(1/v)] v \int_{-1}^1  \frac{dz}{\sqrt{1-z^2}} \left( - \frac{2}{3} \right) v^{-2/3} z^2 (1-z^2)  \nonumber \\
& = & -\frac{1}{3}\frac{ K_\mathrm{Ko2D}^{3/2}}{\nu_{1*}}    \int_0^V dv  [\log(1/v)] v^{1/3} \nonumber \\
& = & -\frac{1}{4}  \frac{ K_\mathrm{Ko2D}^{3/2}}{\nu_{1*}} V^{4/3} \left( \log(1/V) + \frac{3}{4} \right)
\label{eq2D_:Tv_v_1}
\eea
with $\mathrm{Ko2D} = 1.536$ and $\nu_{1*} = 0.1441$. Therefore,
\bea
\int_V^1 dv T(v) & = &  \int_0^1 dv T(v) -\int_0^V dv T(v) \nonumber \\
& = &  -1 +\frac{1}{4}  \frac{ K_\mathrm{Ko2D}^{3/2}}{\nu_{1*}} V^{4/3} \left( \log(1/V) + \frac{3}{4} \right),
\label{eq2D_:Tv_small_v}
\eea
which is plotted in Fig.~\ref{fig:Flux_fraction_CH1} as triangles. Note that the above formula is in good agreement with the numerical results   up to $V \approx  0.2$.  Overall, the consistent picture exhibited in Fig.~\ref{fig:Flux_fraction_CH1} furnish a strong credence to our field-theoretic calculation of energy transfers. 

Let us compare our $\int_V^1 dv T(v)$ with a similar quantity proposed by \cite{Kraichnan:JFM1971_2D3D}, $\int_v^1 ds [Q(s)/s]$, which is the fraction of inertial-range energy transfer from triads for whom the ratio of the smallest to the middle wavenumber is greater than $v$. The two functions are very different, as is evident from Fig.~\ref{fig:Flux_fraction_CH1} and Fig.~2 of \cite{Kraichnan:JFM1971_2D3D}. The most crucial difference is that $\int_0^1 dv T(v) = -1$, but $\int_0^1 ds [Q(s)/s] = 0$. Hence, \cite{Kraichnan:JFM1971_2D3D}'s formula does not capture the inverse energy cascade well.  We believe that the crucial differentiation between \textit{giver} and \textit{receiver} mode in our formalism makes our method more reliable than that of  \cite{Kraichnan:JFM1971_2D3D}.

The fractional energy fluxes discussed above indicate the importance of nonlocal energy transfers in creating the inverse energy cascade in 2D HDT.  In comparison, the local energy transfers (from $p \lessapprox k$) are forward. The opposing nature of local and nonlocal energy transfers is a possible reason for the inadequacy of  RG analysis in providing a correct renormalized viscosity $\nu(k)$ (see Section~\ref{sec:RG_u1}).  Note that 3D HDT too has significant local and nonlocal energy transfers, but they are both forward. Hence, there is no apparent contradiction in the RG  calculation  for 3D HDT (see Section~\ref{sec:ETu2}).

Thus, the field-theoretic calculations of energy transfers provide valuable insights into the turbulence dynamics in 2D HDT, in particular, its inverse energy cascade.  We also remind the reader about the benefits of Craya-Herring basis and the integral $\int dp dz$. Our integrals and the asymptotic analysis are much more compact than those reported earlier using the cartesian basis and $\int dp dq$~\citep{Kraichnan:JFM1959,Orszag:CP1973}.

In the next section, we compute the energy flux for 3D HDT using the same  framework.

\section{Energy Transfers and Flux  for the $u_2$ Component  and 3D HDT}
\label{sec:ETu2}

In 3D HDT, the velocity field is represented by two Craya-Herring components, $u_1$ and $u_2$. In the previous section, we discussed how to compute the energy transfers and flux for the $u_1$ component. In this section, we will perform these calculations for the $u_2$ component.  Note, however, that the net energy transfer in 3D HDT is a sum of the contributions from the $u_1$ and $u_2$ components [see Eq.~(\ref{eq:Suu_kpq_CH})].

\subsection{Mode-to-mode Energy Transfers for the $u_2$ Component } 

As described in Section~\ref{sec:HD_CHbasis}, the mode-to-mode energy transfer from $u_2({\bf p})$ to $u_2({\bf k'})$ with the mediation of $u_1({\bf q})$  is [Eq.~(\ref{eq:Su2u2})] 
\bea
\la S^{u_2 u_2}({\bf k'|p|q}) \ra  & = & - k' \sin\beta \Im \{ \la u_1({\bf q},t) u_2({\bf p},t) u_2({\bf k'},t) \ra \} .\label{eq:Su2u2_avg}
\eea
Note that the energy transfers occurs across the $u_2$ components, but with the mediation  of the $u_1$ component. 

\begin{figure}
	\begin{center}
		\includegraphics[scale = 0.6]{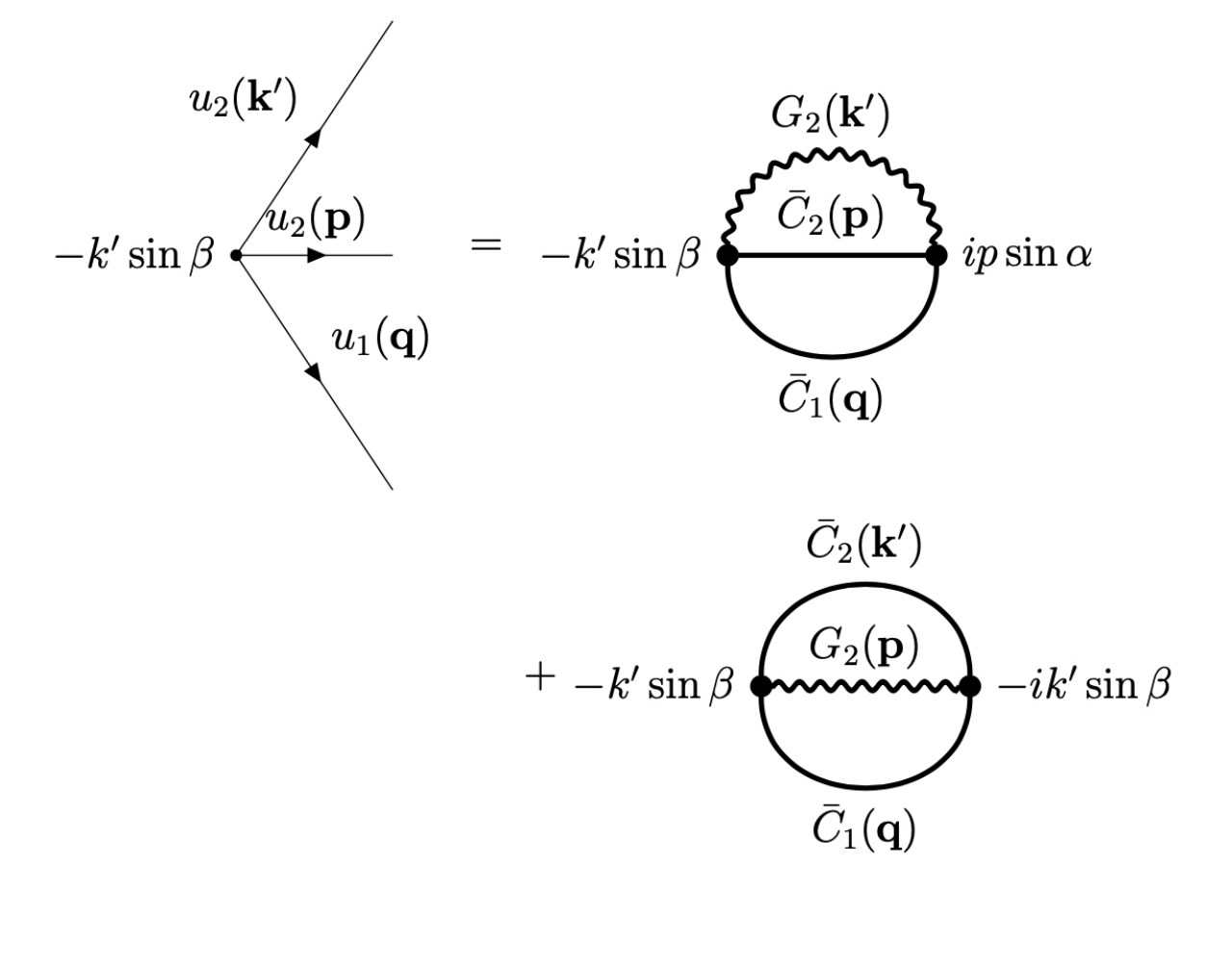}
	\end{center}
	\vspace*{0pt}
	\caption{Feynman diagrams associated with the energy transfers between the $u_2$ components. }
	\label{fig:ET_CH2}
\end{figure}

We employ the  scheme described in Section~\ref{sec:M2M_CH1} to evaluate $\la S^{u_2 u_2}({\bf k'|p|q}) \ra$.  For simplicity, we restrict ourselves to nonhelical flows for which $\la u_1 u_2 \ra =0$. Consequently,   the expansion of $u_2$ components in terms of the Green's function yields nonzero values, whereas the terms arising  from the expansion of $u_1$ component vanishes identically.    The  Feynman diagrams associated with   $\la S^{u_2 u_2}({\bf k'|p|q}) \ra$ are illustrated in Fig.~\ref{fig:ET_CH2}.

Following the same steps as in Section~\ref{sec:M2M_CH1},
the field-theoretic estimate  for  $\la S^{u_2 u_2}({\bf k'|p|q}) \ra $ is
\bea
\la S^{u_2 u_2}({\bf k'|p|q}) \ra  & = & (k \sin\beta)^2   \frac{C_1({\bf q}) [C_2({\bf p}) -C_2({\bf k'})  ] }{\nu(k) k^2 + \nu(p) p^2 + \nu(q) q^2}.
\eea
We assume that turbulence is isotropic, hence $C_1({\bf k'}) = C_2({\bf k'}) = C({\bf k'})$.  In addition, we transform $ \la S^{u_2 u_2}({\bf k'|p|q}) \ra$ to $\la S^{u_2 u_2}({\bf 1|v|w}) \ra$ in terms of variables $u,v$, and $w$ [see Eq.~(\ref{eq:uvw_transform})] as follows:
\bea
\la S^{u_2 u_2}({\bf 1|v|w}) \ra(v,z) & = &     \frac{v^2 w^{-17/3} (v^{-11/3}-1) (1-z^2)}{ (4\pi)^2 (1+v^{2/3} + w^{2/3})}.
\label{eq:SCH2(v,w)}
\eea
\begin{figure}
	\begin{center}
		\includegraphics[scale = 1]{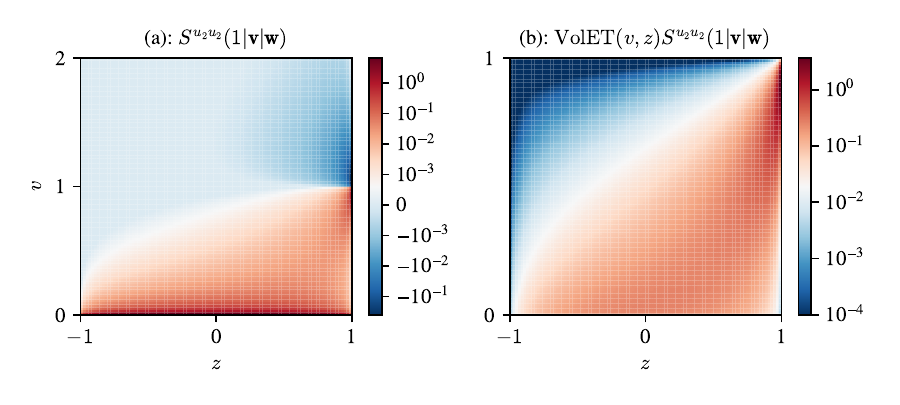}
	\end{center}
	\vspace*{0pt}
	\caption{For energy transfers in 3D HDT: (a) Density plot of $\la S^{u_2 u_2}({\bf k'|p|q}) \ra$. (b)  Density plot of $[v^2 \log(1/v) ] \la S^{u_2 u_2}({\bf k'|p|q}) \ra$.  }
	\label{fig:Skpq_CH2}
\end{figure}
We illustrate $\la S^{u_2 u_2}({\bf 1|v|w}) \ra(v,z)$ in Fig.~\ref{fig:Skpq_CH2}(a).

Equation~(\ref{eq:SCH2(v,w)}) and the density plot of $\la S^{u_2 u_2}({\bf 1|v|w}) \ra(v,z)$  provide the following  insights into the energy transfers for the $u_2$ modes.  Triads with $v \approx w \approx 1$, Case A of Fig.~\ref{fig:triads}, have negligible energy transfer because $(1-v^{-11/3}) \rightarrow 0$ and the other terms of Eq.~(\ref{eq:SCH2(v,w)}) are finite.  For triads with $v \approx w \gg 1$, Case B of Fig.~\ref{fig:triads},  too have negligible energy transfers because
\be
\la S^{u_2 u_2}({\bf 1|v|w}) \ra 
\approx \frac{1}{2 (4\pi)^2} v^2 w^{-17/3} v^{-2/3} (1-z^2)  \rightarrow 0.
\ee

In contrast, as shown in Fig.~\ref{fig:Skpq_CH2},  the energy transfers are significant  (a) when $v \rightarrow 1$  and $z \rightarrow 1$ [centre-right region of Fig.~\ref{fig:Skpq_CH2}(a)]; and (b) when $v \rightarrow 0$ [bottom region of Fig.~\ref{fig:Skpq_CH2}(a)].  Let us examine these two cases using asymptotic analysis.  When $v \rightarrow 1$  and $z \rightarrow 1$ (Cases C and D), $w \rightarrow 0$. Hence,
\bea
\la S^{u_2 u_2}({\bf 1|v|w}) \ra 
& \approx & \frac{1}{(4\pi)^2} \frac{11}{3} (1-z) (1-v) w^{-17/3} \gg 1,
\label{eq:v_to_1}
\eea
which is singular with $\la S^{u_2 u_2}({\bf 1|v|w}) \ra  > 0$ for $v < 1$, but negative for $ v > 1$. These two configurations (Cases C and D) are the primary contributors to the  local energy transfers in 3D HDT.  

Lastly, when   $v \rightarrow 0$ and $w \rightarrow 1$ (Case E), 
\bea
\la S^{u_2 u_2}({\bf 1|v|w}) \ra 
& \approx &  \frac{1}{2 (4\pi)^2} (1-z^2)  v^{-5/3}  \gg 1,
\label{eq:v_ll_1}
\eea
which is also singular.  These triads contribute significantly to the \textit{nonlocal} energy transfers, as described earlier by \cite{Domaradzki:PF1992}, \cite{Verma:Pramana2005S2S}, \cite{Aluie:PF2009}, and others. We will discuss these transfers further in Section~\ref{sec:Pik_3D}.

In Fig.~\ref{fig:Skpq_CH2}(b), we illustrate the density plot of $\mathrm{VolET}(v,z) \la S^{u_2 u_2}({\bf 1|v|w}) \ra $, which is  $v^2 \log(1/v) \la S^{u_2 u_2}({\bf 1|v|w}) \ra$. Note that $\mathrm{VolET}(v,z) \la S^{u_2 u_2}({\bf 1|v|w}) \ra $  is the integrand for the energy flux and shell-to-shell energy transfers, and it has a similar behaviour as $\la S^{u_2 u_2}({\bf 1|v|w}) \ra$.  See Fig.~\ref{fig:Skpq_CH2} for an illustration.
 
In the next subsection, we compute the energy flux for the $u_2$ component.

\subsection{Energy flux for the $u_2$ Component and 3D HDT}
\label{sec:Pik_3D}

In 3D HDT, the energy flux receives contributions from both $u_1$  and $u_2$ components of Craya-Herring basis, as given below:
\be
\la \Pi(R) \ra = \int_{R}^\infty \frac{d{\bf k'}}{(2\pi)^3} \int_0^{R} \frac{d{\bf p}}{(2\pi)^3} [\la S^{u_1 u_1}({\bf k'|p|q}) \ra + \la S^{u_2 u_2}({\bf k'|p|q}) \ra].
\ee
Following the same steps as in Section \ref{sec:Pik_2D} (but for 3D), we deduce that 
\bea
\frac{ \la \Pi(R)\ra}{\epsilon_u }   =  \frac{ K_\mathrm{Ko}^{3/2}}{\nu_{1*}}  
I_4 + \frac{ K_\mathrm{Ko}^{3/2}}{\nu_{2*}}  
I_5 ,
\label{eq:tot_flux_3D_HDT}
\eea
where 
\be
I_4  =\frac{1}{2}   \int_{0.22}^1 dv   [\log(1/v)] v^2 \int_{-1}^1  dz 
\frac{\mathrm{numr}_3}{(1+v^{2/3} + w^{2/3})} ,  
\ee
and
\be
I_5 = \frac{1}{2}  \int_0^1 dv   [\log(1/v)] v^2 \int_{-1}^1  dz  
\frac{v^2 w^{-17/3} (v^{-11/3}-1) (1-z^2)}{  (1+v^{2/3} + w^{2/3})},
\ee
with 
\bea
\mathrm{numr}_3 & = &2 w^{-2} [ (2 v^2z^2 - vz) (vw)^{-11/3} + (v^3 z - 2 v^2 z^2)
w^{-11/3} + (1-v^2)  z v^{-8/3} ]
(1-z^2).
\label{eq:numr_Skpq_u1u1_3D} \nonumber \\
\eea
The prefactor $1/2$ in $I_4$ and $I_5$  arises due to  $8\pi^2/(4 \pi)^2$ [see Eq.~(\ref{eq:Cu_3D}, \ref{eq:dvolET_defn})]. Note that the above 3D integrals are singular. Hence, for better accuracy, we perform the $dz$ integral using the Gauss-Jacobi quadrature, whereas the $dv$ integral using a Romberg iterative scheme.  Refer to Section~\ref{sec:integration} for a brief discussion on the integration procedure.  For the $I_4$ computation, we employ 0.22 as the lower limit for integration that provides better agreement with the 2D computation (see Section~\ref{sec:Pik_2D}).

Our computation yields
\be
I_4 = -0.004954;~~~I_5 = 0.2161.
\ee
We substitute  the above $I_4$ and  $I_5$, as well as  $\nu_{1*} = 0.09218 $  and $\nu_{2*} = 0.4859 $ [Eqs.~(\ref{eq:nu1_3D_integral}, \ref{eq:nu1_for_3D})], in Eq.~(\ref{eq:tot_flux_3D_HDT}), and set $\la \Pi(R) \ra = \epsilon_u$ that yields a numerical value  for the Kolmogorov constant as
\be
K_\mathrm{Ko} = \left( \frac{I_4}{\nu_{1*}} + \frac{I_5}{\nu_{2*}} \right)^{-2/3} = 1.870.
\ee
The above constant is within the range of $K_\mathrm{Ko}$'s reported the past literature, e.g., ~\cite{Kraichnan:JFM1971_2D3D,Leslie:book,Lesieur:book:Turbulence,Nandy:IJMPB1995}.

Interestingly, $I_4 \ll I_5$. Hence, in 3D HDT,  the energy transfers among  $u_1$ components are weaker than those among $u_2$ components. Therefore,  it is reasonable to ignore $u_1$ interactions  for some of the  practical calculations, e.g., for shell-to-shell energy transfers. For such situations, we compute Kolomogrov's constant using
\bea
\frac{ \la \Pi(R)\ra}{\epsilon_u }   =  \frac{ K_\mathrm{KoCH2}^{3/2}}{\nu_{2*}}  
I_5 .
\eea
The above equation yields
\be
K_\mathrm{KoCH2} =  \left(\frac{I_5}{\nu_{2*}} \right)^{-2/3}  = 1.716.
\ee
Thus, we contrast the two constants, $K_\mathrm{KoCH2}$ and $K_\mathrm{Ko}$.  Note that the usage of  Craya-Herring basis provide such fine details on the energy transfers.

\begin{figure}
	\begin{center}
		\includegraphics[scale = 1]{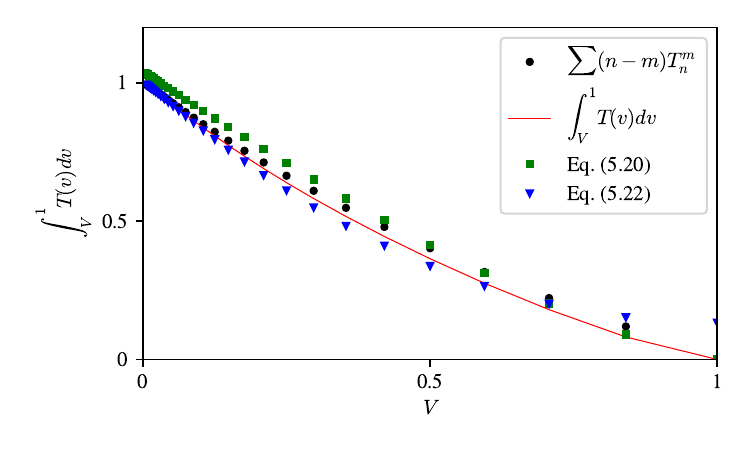}
	\end{center}
	\vspace*{0pt}
	\caption{Plots of the fractional energy flux $\int_V^1 T(v) dv$ (red curve), partial sum of shell-to-shell energy transfers (black dots)  [Eq.~(\ref{eq:cumulativeTmn})], the asymptotic $\int_V^1 T(v) dv$ for small $V$ [Eq.~(5.20)] and for $V \approx 1$ [Eq.~(5.22)]. 	}
	\label{fig:Flux_fraction_CH2}
\end{figure}

For 3D HDT, the fractional energy transfer formula from modes in the wavenumber shell $(R V,R )$ to modes in the wavenumber shell $(R, R V^{-1})$ is  
\bea
\frac{ \la \Pi_V(R)\ra}{\epsilon_u }  &  = & \int_V^1 dv  T(v) \nonumber \\
& = &  \frac{ K_\mathrm{KoCH2}^{3/2}}{\nu_{2*}}  \frac{1}{2}
\int_V^1 dv   [\log(1/v)] v^2 \int_{-1}^1  dz  
\frac{v^2 w^{-17/3} (v^{-11/3}-1) (1-z^2)}{  (1+v^{2/3} + w^{2/3})}.
\label{eq:TkV_3D}
\eea
Refer to Section~\ref{sec:ETu1} and Fig.~\ref{fig:Flux_frac} for the definition.
The  plot of $ \int_V^1 dv  T(v) $, illustrated in Fig.~\ref{fig:Flux_fraction_CH2}, shows that $ \int_V^1 dv  T(v) $  grows with the decrease of $V$. 
Since $\Pi_{V=0}(R) = \Pi(R) = \epsilon$, we deduce that $ \int_0^1 dv  T(v)  = 1$.  A comparsion of Fig.~\ref{fig:Flux_fraction_CH1} and Fig.~\ref{fig:Flux_fraction_CH2} clearly demonstrate that the fractional energy transfers for 3D HDT and 2D HDT are very different.

Now, we estimate $\int_V^1 dv  T(v)$ for $V \approx 1$ using asymptotic analysis.  This quantity represents local energy transfers (see  Fig.~\ref{fig:Skpq_CH2}).  Such transfers are dominated by triads with $v \approx 1$, $z \approx 1$, and $w \approx 0$. Therefore, we make the following change of variables,
\be
v = 1-v';~~~~z = 1-z'.
\ee
In terms of these variables,
\be
w^2 = (1+v^2 -2 v z) \approx (2 z' + v'^2).
\ee
Hence,
\bea
 \int_{-1}^1  dz  
\frac{w^{-17/3}   (1-z^2)}{ (1+v^{2/3} + w^{2/3})}
& \approx &  \int_{0}^1  dz'  (2 z' + v'^2)^{-17/6} 2 z' \nonumber \\
& \approx & \frac{9}{55} v'^{-5/3}.
\label{eq:Int_dz}
\eea
For this particular case, $\la S^{u_2 u_2}({\bf 1|v|w}) \ra 
 \sim  (1-z) (1-v) w^{-17/3}$ is strongly singular [See Eq.~(\ref{eq:v_to_1})].  However, the $dz$ integral gets a dominant contribution from the region near $z \approx 1$ that leads to a taming of the singularity. As derived in Eq.~(\ref{eq:Int_dz}), the integrand after the $dz$ integral is proportional to $v'^{-5/3}$, which is still singular.  Substitution of Eq.~(\ref{eq:Int_dz})   in Eq.~(\ref{eq:TkV_3D}) and the approximation that  $\log(1/v) \approx v'$ yields
\bea
\int_V^1 dv  T(v)  
& = &   \frac{ K_\mathrm{KoCH2}^{3/2}}{\nu_{2*}} \frac{1}{2}  \frac{9}{55}
\int_V^1 dv   [\log(1/v)] v^4  (v^{-11/3}-1)  v'^{-5/3} \nonumber \\
& \approx & \frac{ K_\mathrm{KoCH2}^{3/2}}{\nu_{2*}}   \frac{9}{110} 
\int_0^V dv' v'  \frac{11}{3}  v'  v'^{-5/3} \nonumber \\
& \approx & \frac{ K_\mathrm{KoCH2}^{3/2}}{\nu_{2*}}   \frac{9}{110} \frac{11}{3} 
\int_0^V dv'  v'^{1/3} \nonumber \\
& \approx & \frac{ K_\mathrm{KoCH2}^{3/2}}{\nu_{2*}}   \frac{9}{40} (1-V)^{4/3}.
\label{eq:IntTV_v_1}
\eea

The other limiting case, $V \approx
0$, can be analyzed similarly.  In this case,   $w \approx 1$. Hence,
\bea
\int_0^V dv  T(v)  
& = &  \frac{ K_\mathrm{KoCH2}^{3/2}}{\nu_{2*}}  \frac{1}{2}
\int_0^V dv   [\log(1/v)] v^2 v^2   v^{-11/3} 
\int_{-1}^1 dz  \frac{1}{2} (1-z^2)    \nonumber \\
& \approx & \frac{ K_\mathrm{KoCH2}^{3/2}}{\nu_{2*}} 
\frac{1}{3}  \int_0^V dv [\log(1/v)]  v^{1/3} \nonumber \\
& \approx & \frac{ K_\mathrm{KoCH2}^{3/2}}{\nu_{2*}}
\left( \frac{1}{4}  \log(1/V) + \frac{3}{16}  \right)   V^{4/3}.
\label{eq:IntTV_v_0}
\eea
In the  integral of Eq.~(\ref{eq:IntTV_v_0}), all of $z$'s contribute evenly, unlike the previous case where modes with $z \approx 1$ contribute maximally.  Using $\int_0^1 dv  T(v)  = 1$, we deduce that for small $V$, 
\be
\int_V^1 dv  T(v)  \approx 1- \frac{ K_\mathrm{KoCH2}^{3/2}}{\nu_{2*}}
\left( \frac{1}{4}  \log(1/V) + \frac{3}{16}  \right)   V^{4/3}.
\label{eq:IntTV_smallV_Vto1}
\ee

In Fig.~\ref{fig:Flux_fraction_CH2}, we plot the asymptotic formulas of Eqs.~(\ref{eq:IntTV_v_1}, \ref{eq:IntTV_smallV_Vto1}). We observe that these formula are in good agreement with the exact $\int_V^1 dv  T(v)$ in the respective regimes.   Curiously,  Eqs.~(\ref{eq:IntTV_v_1},  \ref{eq:IntTV_smallV_Vto1})  predict $\int_V^1 dv  T(v)$ for the whole range of $V$ within a factor of 2 or 3. 

Let us physically interpret   Eqs.~(\ref{eq:IntTV_v_1}, \ref{eq:IntTV_v_0}) in some detail.  Note that both the integrals are of the same order, $O(X^{4/3})$, where $X$ is small.  Also note that the relevant integrands ($v^{-5/3}$) and the volumes of the integration ($v^2$) for both the integrals are the same. Hence, we conclude that 
the \textit{cumulative} contributions    from the local triads ($p \lessapprox k$) and the nonlocal ones ($p \ll k$) to the energy flux $\Pi(R)$  are  of the same order.   This observation contradicts the general belief in the \textit{locality of interactions} in HDT.  Earlier, \cite{Verma:Pramana2005S2S} had incorrectly argued that   the cumulative local energy transfer dominates the nonlocal one because the local triads are more in numbers (or have larger volume) than the nonlocal ones.  \cite{Verma:Pramana2005S2S}'s calculations went wrong because of several errors in the complex integrals of the form $\int dp dq$.  We observe that $\int dp dz$ integrals adopted in this paper are more transparent and easier to handle.

We investigate Eqs.~(\ref{eq:TkV_3D}, \ref{eq:IntTV_v_1}, \ref{eq:IntTV_v_0}) further. For small $V$, the wavesphere of radius $RV$ is  tiny. Hence, $p \approx R V$ represents the wavenumber of giver modes. We estimate the wavenumbers of the receiver modes  as $R$. Therefore, $V \approx p/R \approx p/k$, and hence Eq.~(\ref{eq:IntTV_v_0}) yields
\be
\int_0^V dv  T(v)    \approx \frac{ K_\mathrm{KoCH2}^{3/2}}{\nu_{2*}}
\left( \frac{1}{4}  \log\left(\frac{p}{k} \right) + \frac{3}{16}  \right)   \left( \frac{p}{k} \right)^{4/3}.
\label{eq:local_0_V_3d}
\ee
Similarly, for $V \approx 1$, using Eq.~(\ref{eq:IntTV_v_1}) we derive that
\bea
\int_V^1 dv  T(v)  
& \approx & \frac{ K_\mathrm{KoCH2}^{3/2}}{\nu_{2*}}   \frac{9}{40}  \left( \frac{k-p}{k} \right)^{4/3}.
\eea
Since $z \rightarrow 1$ for the type C triads of Fig.~\ref{fig:triads}, we deduce that $k - p \approx q$. Hence,
\bea
\int_V^1 dv  T(v)  
& \approx & \frac{ K_\mathrm{KoCH2}^{3/2}}{\nu_{2*}}   \frac{9}{40}  \left( \frac{q}{k} \right)^{4/3}.
\label{eq:local_V_1_3d}
\eea
Equations~(\ref{eq:local_0_V_3d}, \ref{eq:local_V_1_3d}) indicate that the nonlinear energy transfers  decrease as $p/k$ or $q/k$ decrease. This phenomena is termed as \textit{local interactions}.  Many authors, e.g., \cite{Kraichnan:JFM1971_2D3D}, \cite{Domaradzki:PF1990}, \cite{Zhou:PF1993}, \cite{Verma:Pramana2005S2S},  \cite{Aluie:PF2009},  arrived at similar conclusions.   In contrast, using $k^{-3/2}$ energy spectrum, \cite{Kraichnan:JFM1959} predicted $\int_V^1 dv  T(v)   \sim (p/k)^{3/2}$, which is incorrect. Note, however, that 3D HDT has  significant energy transfers from nonlocal triads, as described above. 

It is instructive to  study $\int_V^1 dv T(v)$ for some sample  $V$'s. We observe that
$\int_{0.7}^1 dv T(v) \approx 0.18$.  Hence, only 18\% energy is transferred from the giver modes with $ v \ge 0.7$.  Thus, only a  fraction of energy is transferred from triads with $k \approx p$, in contrast to the general belief  in local energy transfers in 3D HDT. Also, $\int_{0.1}^1 dv T(v) \approx 0.83$. Consequently, around 17\% contribution to the energy flux comes from  the nonlocal triads for which the giver wavenumbers have $v < 0.1$, consistent with the above discussion.

In summary, our calculations on the mode-to-mode energy transfers and energy flux, as well as their asymptotic limits, show that in 3D HDT, the cumulative local and nonlocal energy transfers are of the same order, and they are forward. The foward nonlocal energy transfers do not cancel the local transfers, which are encountered in the RG analysis.  This is possibly the reason why  the RG predictions of $\nu^{(n)}_2$   for 3D HDT are similar. Note, however, that $\nu^{(n)}_1$  predictions for 2D HDT have significant variations including the sign changes. We attribute these contradictory results to the   nonlocal energy transfers that oppose the local  transfers in 2D HDT.

The above detailed calculations exhibit the complexity of  energy transfers  in HDT.  In the next two sections we report the shell-to-shell energy transfers in 3D and 2D HDT.

\section{Shell-to-shell Energy Transfers in 3D HDT}
\label{sec:S2S_3D}

In comparison to the energy flux, shell-to-shell (S2S) energy transfers provide a more detailed information about the turbulence dynamics. EDQNM calculations and numerical simulations reveal that maximal energy transfers take place among the neighbouring shells, and that the energy transfer decreases rapidly when the distance between the giver and receiver shells increases~\citep{Domaradzki:PF1990,Verma:Pramana2005S2S,Aluie:PF2009}.   \cite{Verma:Pramana2005S2S}  performed a preliminary calculation of the S2S energy transfers using field theory. The present field-theoretic computations  expand the previous work, as well as correct some errors in them.

We divide the wavenumber space logarithmically into shells~\citep{Domaradzki:PF1990,Dar:PD2001,Lesieur:book:Turbulence}, with the $n$-th shell representing the wavenumbers  band  $(k_0 s^n, k_0 s^{n+1})$.  In this paper, we compute the shell-to-shell energy transfer from the $m$-th shell to the $n$-th shell using
\be
\la T^m_n   \ra = \int_{k_0 s^n}^{k_0 s^{n+1}} \frac{d{\bf k'}}{(2\pi)^3}   \int_{k_0 s^m}^{k_0 s^{m+1}} \frac{d{\bf p}}{(2\pi)^3}    \la S^{u_2 u_2}({\bf k'|p|q}) \ra.
\ee
The above formula, proposed by \cite{Dar:PD2001} and \citet{Verma:PR2004},  removes the ambiguity between the giver and receiver wavenumbers that plagues the earlier works~\citep{Domaradzki:PF1990}. In our S2S energy transfer computation for 3D HDT, we ignore the energy transfers via the $u_1$ channel because they are weaker than those via the $u_2$ channel. 

In the following discussion, we study the S2S energy transfers in the inertial range, which is scale invariant. For these computations, it is convenient to  employ the following change of variables for a triad $(k,p,q)$~\citep{Verma:Pramana2005S2S}:
\be
k = \frac{k_0 s^n}{u};~~~p = \frac{k_0 s^n}{u} v ;~~~q = \frac{k_0 s^n}{u} w.
\ee
In terms of these variables,
\bea
\frac{ \la T^m_n \ra}{\epsilon_u }   =   \frac{ K_\mathrm{KoCH2}^{3/2}}{\nu_{2*}}  \frac{1}{2}
\int_{1/s}^1 \frac{du}{u} \int_{u s^{m-n}}^{u s^{m-n+1}} dv    v^4  (v^{-11/3}-1) \int_{-1}^1  dz  
\frac{w^{-17/3}  (1-z^2)}{  (1+v^{2/3} + w^{2/3})}.
\eea
Since we ignore the contributions from the $u_1$ component of the mode, we employ $ K_\mathrm{KoCH2} $ and $ \nu_{2*} $ for the above computation.

Note that $T^m_n$ is a function of $m-n$, which is a consequence of scale invariance of isotropic and homogeneous  HDT. Also, $T^m_m = 0$, and $T^m_n = -T^n_m$~\citep{Dar:PD2001,Teaca:PRE2009}, which  follows from the following property of mode-to-mode energy transfers~\citep{Dar:PD2001},
\be
S^{uu} ({\bf k'|p|q}) = - S^{uu} ({\bf p|k'|q}) .
\ee

\begin{figure}
	\begin{center}
		\includegraphics[scale = 0.8]{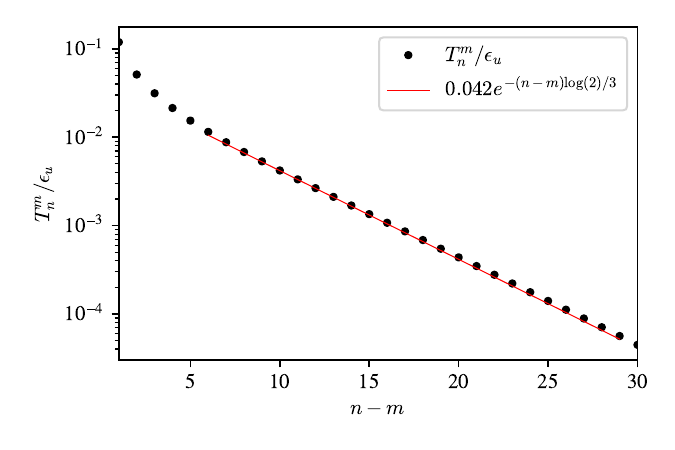}
	\end{center}
	\vspace*{0pt}
	\caption{For 3D HDT, plot of the normalized shell-to-shell energy transfer, $T^m_n/\epsilon_u$, vs. $n-m$. For large $n-m$, the best fit curve to $T^m_n/\epsilon_u$ is $0.042 \exp[-(n-m) \log(2)/3]$.	}
	\label{fig:Tmn_CH2}
\end{figure}

For our analysis, we take $s = 2^{1/4}$ and $n-m$ ranging from 1 to 30. For these parameters, we compute the above integral numerically using the Gauss-Jacobi quadrature for the $dz$ integral, and  Romberg schemes for the $dv$ and $du$ integrals. In Fig.~\ref{fig:Tmn_CH2}, we plot the computed $T^m_n$'s   as a function of $n-m$. We observe that for $n-m \gg 1$, 
\be
T^m_n \approx 0.042 \exp[-(n-m) \log(2)/3],
\ee
which can be explained as follows.

When $n-m \gg 1$, the magnitudes of the giver wavenumbers are much smaller  than those of receiver ones, i.e., $v \ll 1$. Hence, the derivation of $T^m_n$ with $n -m \gg 1$ follows similar steps as that of Eq. (\ref{eq:IntTV_v_0}). Therefore, using $K_\mathrm{KoCH2} = 1.716$ and $\nu_{2*} = 0.4859$, we obtain
\bea
\frac{ \la T^m_n \ra}{\epsilon_u }   & \approx &    \frac{ K_\mathrm{KoCH2}^{3/2}}{\nu_{2*}}  
\int_{1/s}^1 \frac{du}{u} \int_{u s^{m-n}}^{u s^{m-n+1}} dv    v^4  v^{-11/3} \int_{-1}^1  dz  
\frac{(1-z^2)}{ 4} \nonumber \\
& \approx & \frac{ K_\mathrm{KoCH2}^{3/2}}{\nu_{2*}}  
\int_{1/s}^1 \frac{du}{u} \int_{u s^{m-n}}^{u s^{m-n+1}} dv  \frac{1}{3}   v^{1/3} \nonumber \\
& \approx & \frac{ K_\mathrm{KoCH2}^{3/2}}{\nu_{2*}}   
\frac{3}{16} \frac{(s^{4/3}-1)^2}{s^{4/3}} s^{-4 (n-m)/3}  \nonumber \\
& \approx & \frac{ K_\mathrm{KoCH2}^{3/2}}{\nu_{2*}}   
\frac{3}{16} \frac{(s^{4/3}-1)^2}{s^{4/3}}  
\left(  \frac{K}{P} \right)^{-4/3}  \nonumber \\
& \approx & 0.0465 e^{-(n-m) \log(2)/3},
\label{eq:S2S_3d}
\eea
which is quite close to the best-fit curve, $0.042 e^{-(n-m) \log(2)/3}$ (see Fig.~\ref{fig:Tmn_CH2}).  In the second-last line of Eq.~(\ref{eq:S2S_3d}), $K=k_0 s^n$ and $P=k_0 s^m$ represent the receiver and giver wavenumbers respectively.  Thus, the above asymptotic formulas provide a reasonably accurate estimation of $T^m_n$'s for large $n-m$s.

Equation~(\ref{eq:S2S_3d})  shows that for distant shells, the S2S energy transfer decreases as $(K/P)^{-4/3}$.  This sharp decrease in the S2S energy transfers  demonstrates the locality of the energy transfers in 3D HDT, as discussed  in the past literature. For example, \cite{Zhou:PF1993} showed that the fractional energy flux $\Pi(k,s) \sim s^{-4/3}$, where $s(k,p,q) = \mathrm{max}(k,p,q)/\mathrm{min}(k,p,q)$.  These results are consistent with those on the partial energy flux presented in Section~\ref{sec:ETu2}.

%We can also derive the $T^1_2 \approx 0.1395$, which is close to the exact result of 0.1193. 

Before closing this section, we remark that $\int_V^1 dv T(v)$  is related to the following cumulative S2S energy transfers:
\bea 
\int_{V = 2^{-M}}^1 dv T(v) & \approx & \sum_{n-m = 1}^{M} (n-m) T^m_n ,
\label{eq:cumulativeTmn} \\
\int_{0}^{V = 2^{-M}} dv T(v) & \approx & \sum_{n-m = M}^{\infty} (n-m) T^m_n .
\label{eq:cumulativeTmn2}
\eea
In Fig.~\ref{fig:Flux_fraction_CH2}, we plot  the quantities in the left-hand-side and right-hand-side of Eq.~(\ref{eq:cumulativeTmn})  and observe them to be in close agreement  with each other. In addition, for large $M$, using variables $x =n-m$ and $a = \log(2)/3$, as well as Eq.~(\ref{eq:S2S_3d}), we obtain
\bea 
\sum_{n-m = M}^{\infty} (n-m) T^m_n 
& \approx & \int_M^\infty 0.0465 x \exp(-ax) 
\nonumber \\
& \approx &  0.0465 \frac{M}{a} 
  \exp(-aM ) \nonumber \\
& \approx &  1.156 
  \left( \frac{K}{P} \right)^{-4/3}  \log\left( \frac{K}{P} \right) ,
  \label{eq:Tmn_Mtoinfty}
\eea
where $s^M = K/P$.   Equation~(\ref{eq:Tmn_Mtoinfty}) is the right-hand-side of Eq.~(\ref{eq:cumulativeTmn2}), and it is comparable to its left-hand-side, which is [see Eq.~(\ref{eq:IntTV_v_0})]
\be
\int_0^V dv T(v) \approx  1.515  \left(  \log\left( \frac{K}{P} \right) + \frac{3}{4} \right)
\left( \frac{K}{P} \right)^{-4/3} .
\ee
 These results show that   the cumulative S2S transfers are in agreement with the fractional energy fluxes computed in Section~\ref{sec:ETu2}, and they are  $O((K/P)^{-4/3})$.
 
 In summary, for 3D HDT, the S2S energy transfers decreases as the distance between the giver and receiver shell increases. The energy transfers between the distant shells is rather small. However, many of these transfers add up to a significant nonlocal energy transfers, and they  are comparable to the local  transfers. 
 
 In the next section, we will compute the S2S energy transfers for 2D HDT using field theory.

\section{Shell-to-shell Energy Transfers in 2D HDT}
\label{sec:S2S_2D}

In this section, we compute S2S energy transfers for 2D HDT and show that the local energy transfers are forward, whereas the nonlocal ones are backward.

In 2D HDT, the shell-to-shell energy transfers take place between the $u_1$ components.  Hence,
\bea
\la T^m_n   \ra & = & \int_{k_0 s^n}^{k_0 s^{n+1}}   \frac{d{\bf k'}}{(2\pi)^2}   \int_{k_0 s^m}^{k_0 s^{m+1}}   \frac{d{\bf p}}{(2\pi)^2}    \la S^{u_1 u_1}({\bf k'|p|q}) \ra. 
\eea
Following similar steps as in Section~\ref{sec:S2S_3D} (for 3D HDT), we obtain
\bea
\frac{ \la T^m_n \ra}{\epsilon_u }   =   \frac{ K_\mathrm{Ko2D}^{3/2}}{\nu_{1*}}  \frac{4}{\pi}
\int_{1/s}^1 \frac{du}{u} \int_{u s^{m-n}}^{u s^{m-n+1}} dv    v  \int_{-1}^1  \frac{dz}{ \sqrt{1-z^2}}  
\frac{\mathrm{numr}_2}{(1+v^{2/3} + w^{2/3}) } .
\label{eq:S2S_2D}
\eea
 We employ $K_\mathrm{Ko2D} = 1.536$ and $\nu_{1*} = 0.1441$ for the shell-to-shell computation for 2D HDT.

\begin{figure}
	\begin{center}
		\includegraphics[scale = 0.8]{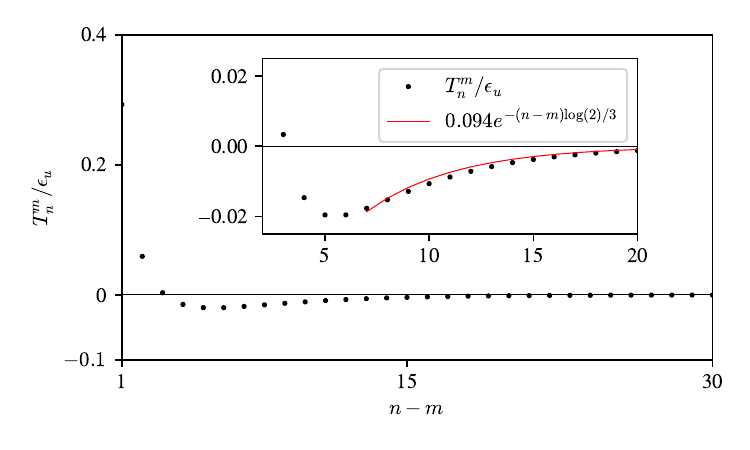}
	\end{center}
	\vspace*{0pt}
	\caption{For 2D HDT, the normalized shell-to-shell energy transfer, $T^m_n/\epsilon_u$, vs. $n-m$.  The curve exhibits forward energy transfers for the local shells ($n-m= 1,2,3$), but inverse energy transfers for the distant shells. In the inset, for large $n-m$, the best fit curve to $T^m_n/\epsilon_u$ is $-0.094 \exp[-(n-m) \log(2)/3]$.}
	\label{fig:Tmn_CH1}
\end{figure}

Using Eq.~(\ref{eq:S2S_2D}), we compute $\la T^m_n \ra/\epsilon_u $ for 2D HDT with  $s = 2^{1/4}$ and $n-m$ ranging from 1 to 30. These results, plotted in Fig.~\ref{fig:Tmn_CH1}, exhibit both forward and inverse  S2S energy transfers. Note that $T^m_{m+1}$, $T^m_{m+2}$, $T^m_{m+3}$ are positive, while the rest of $T^m_n$'s are negative. The latter energy transfers contribute to the inverse energy cascade. In addition,  we compute the quantities on both  sides of Eq.~(\ref{eq:cumulativeTmn}).  As shown in Fig.~\ref{fig:Flux_fraction_CH1}, these quantities are nearly equal, showing consistency between the  S2S energy transfers and the fractional energy fluxes, $\int_V^1 dv T(v)$.

For large $n-m$, 
\be
\frac{ \la T^m_n \ra}{\epsilon_u }   \approx -0.094 \exp[-(n-m) \log(2)/3],
\label{eq:Tmn_2D_fit}
\ee
similar to that for 3D HDT, except the negative sign that indicates an inverse energy transfers for distant or \textit{nonlocal } shells.   We can derive the above equation using an asymptotic approximation under the limit $v \ll 1$. 
We approximate Eq.~(\ref{eq:S2S_2D}) using the approximation of Eq.~(\ref{eq:2D_v_ll_1}) and obtain

\bea
\frac{ \la T^m_n \ra}{\epsilon_u }   & \approx &    \frac{ K_\mathrm{Ko2D}^{3/2}}{\nu_{1*}}  \frac{4}{\pi}
\int_{1/s}^1 \frac{du}{u} \int_{u s^{m-n}}^{u s^{m-n+1}} dv    v v^{-2/3} \int_{-1}^1   \frac{dz}{\sqrt{1-z^2}} \left( - \frac{2}{3} \right)  z^2 (1-z^2)
 \nonumber \\
& \approx &-\frac{1}{3}   \frac{ K_\mathrm{Ko2D}^{3/2}}{\nu_{1*}}  
\int_{1/s}^1 \frac{du}{u} \int_{u s^{m-n}}^{u s^{m-n+1}} dv    v^{1/3} \nonumber \\
& \approx & -\frac{3}{16} \frac{ K_\mathrm{Ko2D}^{3/2}}{\nu_{1*}}   
\frac{(s^{4/3}-1)^2}{s^{4/3}} s^{-4 (n-m)/3} \nonumber \\
& \approx & -\frac{3}{16} \frac{ K_\mathrm{Ko2D}^{3/2}}{\nu_{1*}}   
\frac{(s^{4/3}-1)^2}{s^{4/3}} 
\left(  \frac{K}{P} \right)^{-4/3} \nonumber \\
& \approx & -0.13 e^{-(n-m) \log(2)/3},
\label{eq:Tmn_asymptote_kbyp}
\eea
where $K=R s^n$ and $P=R s^m$ represent the receiver and giver wavenumbers respectively.
Note the above prefactor of $-0.13$ is of the same order as the prefactor of Eq.~(\ref{eq:Tmn_2D_fit}), which is remarkable considering major approximations made in the asymptotic analysis.  Equation~(\ref{eq:Tmn_asymptote_kbyp}) indicates that the energy transfer decays as $(K/P)^{-4/3}$, similar to that in 3D HDT.

In summary, in 2D HDT, the S2S energy transfers among the neighbouring inertial shells are forward, but those between the distant shells are small but backward. However, latter transfers add up to yield an inverse constant energy cascade ($\Pi(R) = -\epsilon_u$).  These results are consistent with the fractional energy flux analysis of Section~\ref{sec:ETu1}.

We conclude in the next section.

\section{Conclusions}

In this paper,  we employ perturbative field theory to  incompressible Navier-Stokes equation in 2D and 3D, and compute the renormalized viscosities, energy transfers, energy fluxes, and shell-to-shell energy transfers. The novelty of our approach is an application of Craya-Herring basis that enables separate evaluation of the renormalized viscosities and energy transfers for the two components, $u_1$ and $u_2$. These novel calculations  provide many valuable insights into the turbulence dynamics.

For 2D HDT, we successfully compute the mode-to-mode energy transfers, energy flux, and shell-to-shell energy transfers. These computations reveal forward energy transfer for neighbouring wavenumbers or shells, and inverse energy transfer for the distant wavenumbers or shells.  That is, in a triad  $(k,p,q)$,  where $k, p, q$ are the receiver, giver, and mediator wavenumbers respectively, the energy flows from  $p$ to $k$ when $p \lessapprox k$ (local), but from $k$ to $p$ when  $p \ll k$ (nonlocal).  In addition, in 2D HDT,  individual local transfers are significant, but the nonlocal ones are weak and decrease as $(k/p)^{-4/3}$.  However, the cumulative nonlocal energy transfer is significantly large and it overcompensates the forward local energy transfers to yield an inverse energy cascade. 

For 2D HDT, the renormalized viscosity $\nu^{(n)}_1$ computed using the RG analysis is somewhat uncertain.  Our $\nu^{(n)}_1$ is positive, but it is around 4 times smaller than those reported by \cite{Kraichnan:JFM1971_2D3D}, \cite{Olla:PRL1991}, and \cite{Nandy:IJMPB1995}. Note, however, that some researchers reported negative $\nu^{(n)}_1$, while \cite{Liang:PF1993} argued that the RG fixed point for 2D HDT is unstable.   We believe that the inconsistencies in $\nu^{(n)}_1$ is due to the inability of the RG analysis to take into account the distant or nonlocal interactions.  The opposing nature of the local and nonlocal interactions in 2D HDT makes the RG analysis quite complex, and it needs to be refined in future.  

Interestingly, the eddy viscosity for 2D HDT is negative because it takes into account both local and nonlocal energy transfers (see Appendix \ref{sec:eddy_visc}). On the contrary, the positive $\nu^{(n)}_1$ computed by some authors is due to the forward energy transfers between the local wavenumbers, which goes into the RG calculations. Note that the sweeping effect~\citep{Kraichnan:PF1964Eulerian} is related to the nonlocal interactions between the large-scale structures and small-scale fluctuations. Hence, ours and Kraichnan's arguments on the inapplicability of field-theoretic tools to HDT in Eulerian framework  have certain similarities. 

In 3D HDT, each velocity mode has two components in  Craya-Herring basis. In this paper, we  compute the 
renormalized viscosities ($\nu^{(n)}_1$ and $\nu^{(n)}_2$) and energy transfers for both the components. Interestingly, $\nu^{(n)}_1 \ll \nu^{(n)}_2$, and the mode-to-mode energy transfers and the energy flux via  $u_1$ channel is much weaker than those via  $u_2$ channel. Hence, we focus on the latter energy transfers.  

For 3D HDT, we compute the shell-to-shell energy transfers using field theory and find both local and nonlocal energy transfers to be forward.  The local transfers are strong, but the nonlocal ones are weak and they fall off as $(k/p)^{-4/3}$. These results are consistent with the earlier works, e.g., \cite{Domaradzki:PF1990}, \cite{Verma:Pramana2005S2S}, and \cite{Aluie:PF2009}.   More importantly, we compute the cumulative local and nonlocal energy transfers via fractional energy flux and shell-to-shell energy transfers. We show  that \textit{the cumulative local and  nonlocal energy transfers are of the same order}.  This new result is contrary to earlier predictions by \cite{Domaradzki:PF1990}, \cite{Verma:Pramana2005S2S}, and \cite{Aluie:PF2009} that the cumulative local energy transfers dominate the nonlocal ones.  Earlier, \cite{Verma:JPA2007} had argued that the origin of the \textit{bottleneck effect} may be the nonlocal energy transfers.  There are possibly more consequences of our finding on turbulence modelling.

A typical  derivation of $k^{-5/3}$ energy spectrum for HDT employs   locality of interactions. Does the nonlocal energy transfers discussed in this paper spoil this scaling argument? Fortunately, this is not the case because  cumulative energy transfers for both local and nonlocal triads  scale as $(k/p)^{-4/3}$ [see Eqs.~(\ref{eq:IntTV_v_0}, \ref{eq:IntTV_v_1})];  this common scaling law for the energy transfers saves the scaling arguments employed in the derivation of $k^{-5/3}$ spectrum.

For 3D HDT, our RG predictions of $\nu^{(n)}_2$ and the Kolmogorov constant are in agreement with the past results by various authors~\citep{Yakhot:JSC1986,McComb:book:Turbulence,Zhou:PR2010}.  Note that the coarse-graining in RG analysis takes into account the local interactions, not the nonlocal ones. Since the forward nonlocal energy transfers do not counter the local ones in 3D HDT,   $\nu^{(n)}_2$ computed using the RG analysis do not vary significantly. This feature differs from    $\nu^{(n)}_1$ of 2D HDT that exhibits significant variations because of the opposing directions of energy transfers for the local and nonlocal triads.

In summary, our field-theoretic calculations provide many valuable insights into the turbulence dynamics, in particular on the local and nonlocal energy transfers.  In this paper we report the renormalized viscosities and the Kolmogorov constants for 2D and 3D HDT. It is utmost important to test these predictions using high-resolution numerical simulations and experiments.  The renormalized viscosity can be computed using the two-time correlation functions in the inertial range~\citep{Verma:INAE2020_sweeping,Verma:PRE2023_shell}, whereas the fractional energy flux and shell-to-shell energy transfers can be computed using Fourier transforms~\citep{Dar:PD2001,Verma:book:ET}. Note, however, that these computations may require very high-resolution (e.g., $8192^3$ grid) and long numerical simulations.   We hope that these tests will be performed in near future.

The framework presented in this paper is  general that may be applicable to other flows, e.g., scalar turbulence, magnetohydrodynamic turbulence, rotating turbulence, etc.  We plan to perform these calculations in future.

\vspace{1cm} 
The author thanks Soumyadeep Chatterjee for help in generating the Feynman diagrams and Figure 18 of the paper. He also thanks Rodion  Stepanov, Jayant Bhattacharjee, and Anurag Gupta for suggestions.  This work is supported by  Projects  SERB /PHY /20215225 and SERB /PHY /2021473 by the Science and Engineering Research Board, India.

%%%%
\appendix

\section{Evaluation of the Renormalization and Energy-Flux Integrals}
\label{sec:integration}

Computation of field-theoretic  integrals, which are typically  singular,  is always tricky~\citep{Peskin:book:QFT}.  In this section, we briefly describe how we overcome singularities in our calculations.

For the RG procedure, the integrals of  Eqs.~(\ref{eq:nu1_integral}, \ref{eq:nu2_integral}) are finite because they are performed in the band $1 \le p' \le b$ and $1 \le q' \le b$. Here, we employ the Gaussian quadrature for the $dq'$ integral and a Romberg scheme for the $dp'$ integral. This procedure yields finite and accurate results. However, the integrals for the energy fluxes are singular and they need special attention.

In 2D HDT, the  energy-flux integral is  of the following form:
\bea
I =  
\int_0^1 dv   [\log(1/v)] v \int_{-1}^1  dz    f(v,z) \sqrt{1-z^2} ,
\label{eq:Pi_2D_integral_proc}
\eea
where $ f(v,z)$ involves singularities. For accurate evaluation of $dz$ integration, we employ the Gauss-Jacobi quadrature:
\be
\int_{-1}^1  dz    f(v,z)  (1-z)^{1/2} (1+z)^{1/2}
\approx \sum_k f(v,z_k) w_k,
\ee
where $z_k$ is the $k$th root of Jacobi polynomials, and $w_k$ is the corresponding \textit{weight}. Note that   $f(v,z_k)$ is evaluated  at $z = z_k$. The Gauss-Jacobi quadrature yields finite answer for these singular integrals.  We employ a Romberg iterative scheme for the subsequent $dv$ integration.  

In 3D HDT, the  energy-flux integral is  of the following form:
\bea
I =   
\int_0^1 dv   [\log(1/v)] v^4 \int_{-1}^1  dz    g(v,z) (1-z^2) .
\label{eq:Pi_3D_integral_proc}
\eea
The $dz$ integration for the above form is performed using the Gauss-Jacobi quadrature as follow:
\be
\int_{-1}^1  dz    g(v,z)  (1-z)^{1} (1+z)^{1}
= \sum_k g(v,z_k) w'_k,
\ee
where $\{ z_k \}$ and $\{ w'_k \}$ are the roots and weights of the Jacobi polynomials. After this, a Romberg iterative scheme is employed for the  $dv$ integration.  

The integral computations for the shell-to-shell energy transfers  are similar to those for the energy fluxes.

\section{Eddy Viscosity and its Relation to the Renormalized Viscosity}
\label{sec:eddy_visc}

Eddy viscosity is an important quantity in atmospheric physics and astrophysics. It represents an enhanced viscosity at large scales due to turbulence. \cite{Kraichnan:JAS1976} and \cite{Leith:JAS1972}  computed the eddy viscosity using EDQNM approximation and field theory. 

\cite{McComb:PRA1983} and \cite{Zhou:PRA1989} argued that eddy viscosity is same as the renormalized viscosity. In fact, they use the phrase ``renormalized eddy viscosity" for $\nu^{(n)}$. In this Appendix, we relate these two viscosities using the RG equation.

The RG equation for velocity field is~\citep{Yakhot:JSC1986}
\bea
[\partial_t +\nu^{(n)}  k^2 ]{\bf u}^<({\bf k'},t) & = & -i  \int \frac{d{\bf p}} {(2\pi)^{d}}   [{\bf k \cdot u^<(q},t)] {\bf u^<(p},t) + {\bf F^<(k'},t)  ,
\label{eq:uk_RG} 
\eea
where $\nu^{(n)} $ is the renormalized viscosity at $k=k_n$, the cutoff wavenumber;  ${\bf F^<(k'},t)$ is the external force; and ${\bf k'+p+q} =0$.  In Eq.~(\ref{eq:uk_RG}), the  convolution $\int d{\bf p}$ is performed over $0 < p \le k_n$ (also see Section~\ref{sec:RG}).  Here, we employ cartesian coordinate system that simplifies the derivation of the  connection between the renormalized viscosity and eddy viscosity. Using Eq.~(\ref{eq:uk_RG}), we derive the following equation for the modal energy $E(k',t) = |{\bf u}^<({\bf k'},t)|^2/2$:
\bea
[\partial_t + 2 \nu^{(n)}   k^2 ] E^<({\bf k'},t)& = &   \int \frac{d{\bf p}} {(2\pi)^{d}}   \Im \{ [{\bf k \cdot u^<(q},t)] [{\bf u^<(p},t) \cdot {\bf u^<(k'},t)  ] \} \nonumber \\ 
&& + \Re [  {\bf F^<(k'},t)  \cdot  {\bf u^<(k'},t) ],
\label{eq:Ek_RG} 
\eea
where $\Re[.]$ and $ \Im[.]$ represent, respectively, the real and imaginary parts of the argument.  Note that
\be 
\int \frac{d{\bf p}} {(2\pi)^{d}}   \Im \{ [{\bf k \cdot u^<(q},t)] [{\bf u^<(p},t) \cdot {\bf u^<(k'},t)  ] \} = 0
\label{eq:Ek_eddy_visc_tot}
 \ee
 because of the \textit{ detailed conservation of energy}~\citep{Kraichnan:JFM1959,Verma:book:ET}. The above identity also follows from the fact that 
 \be
\int d{\bf x} [({\bf u \cdot \nabla u}) \cdot {\bf u}] = 
\int d{\bf x} \nabla \cdot [\frac{u^2}{2} {\bf u}] = 
0,
\ee
where the derivatives are performed at length scale $2\pi/k_n$. As a consequence of the above conservation law, Eq.~(\ref{eq:Ek_eddy_visc_tot}) gets simplified to
\bea
\partial_t  E^<({\bf k'},t)= -2 \nu^{(n)}   k^2 E^<({\bf k'},t) + \Re [  {\bf F^<(k'},t)  \cdot  {\bf u^<(k'},t) ].
\label{eq:eddy_visc_defn} 
\eea
Equation~(\ref{eq:eddy_visc_defn}) has the same structure as the dissipative Navier-Stokes equation, except that $\nu$ of the original equation is replaced by $\nu^{(n)} $.   The renormalized viscosity varies with  wavenumber, hence it is called \textit{eddy viscosity}. 

We can derive interesting relations based on Eq.~(\ref{eq:eddy_visc_defn}). An integration Eq.~(\ref{eq:eddy_visc_defn}) from $k=0$ to $k_n$ yields
\bea
\partial_t  \int_0^{k_n} dk' E^<({\bf k'},t)= -\int_0^{k_n} dk' 2 \nu^{(n)}   k^2 E^<({\bf k'},t) + \int_0^{k_n} dk' \Re [  {\bf F^<(k'},t)  \cdot  {\bf u^<(k'},t) ].  \nonumber \\
\label{eq:energy_integrate_eddy}
\eea
For 3D HDT, under steady state ($\partial_t \int_0^{k_n} [.] = 0$). Hence,
\be
\int_0^{k_n} dk' 2 \nu^{(n)}   k^2 E^<({\bf k'},t) = \int_0^{k_n} dk' \Re [  {\bf F^<(k'},t)  \cdot  {\bf u^<(k'},t) ] = \epsilon_u.
\label{eq:eddy_visc_3d_integ}
\ee
That is, the energy injected at large scales leaves the  wavenumber sphere as the energy flux. This feature forms a basis for the \textit{Large Eddy Simulations}. Interestingly, with $E(k,t) = K_\mathrm{Ko} \epsilon_u^{2/3} k^{-5/3}$, 
the integral of Eq.~(\ref{eq:eddy_visc_3d_integ}) yields
\be
\int_0^{k_n} dk' 2 \nu^{(n)}   k^2 E^<({\bf k'},t) =
\frac{3}{2} \nu_{2*} K_\mathrm{KoCH2}^{3/2} \epsilon_u 
\approx 1.638 \epsilon_u,
\ee
which is close to $\epsilon_u$. This is reasonable considering various approximations made in the calculation, e.g., $k^{-5/3}$ extends to all the wavenumber range.

Equation~(\ref{eq:energy_integrate_eddy}) leads to some inconsistencies in 2D HDT that is forced at $k=k_f$.  For $k < k_f$, in the absence of ${\bf F^<(k'},t)$, Eq.~(\ref{eq:energy_integrate_eddy}) predicts that 
\be
\partial_t  \int_0^{k_n} dk' E^<({\bf k'},t) = 
-\int_0^{k_n} dk' 2  \nu^{(n)} k^2 E^<({\bf k'},t)  < 0
\ee
for positive $ \nu^{(n)}$. That is, the large-scale energy  decays with time in spite of the inverse energy cascade.  This is a contradiction that can be resolved only if $ \nu^{(n)} < 0$. Note that \cite{Kraichnan:JAS1976} predicted  eddy viscosity to be negative for 2D HDT based on the energy flux computation. On the contrary, RG analysis yields positive $ \nu^{(n)}$ due to the forward energy transfers for the local interactions (see Section~\ref{sec:RG_u1}).  These observations indicate that we need a closer look at the field-theoretic treatment of 2D HDT, as well as local interaction issues in the RG treatment of turbulence.
 
\section{Enstrophy Transfers in 2D HDT}
\label{sec:Enstrophy}

In a  2D HDT that is forced at $k=k_f$, we observe a constant energy flux and $k^{-5/3}$ energy spectrum for $k < k_f$, and a constant enstrophy flux and $k^{-3}$ energy spectrum for $k>k_f$~\citep{Kraichnan:PF1967_2D, Boffetta:ARFM2012}. In Sections~\ref{sec:ETu1} we performed field-theoretic computation of energy transfers for $k < k_f$ regime.    In this Appendix, we  study the nature of enstrophy transfers and enstrophy flux in the $k < k_f $ regime.

\subsection{Mode-to-mode Enstrophy Transfer} 

For 2D HDT, the enstrophy transfer rate from $\omega({\bf p})$ to $\omega({\bf k})$ with the mediation of ${\bf u}({\bf q})$ is~\citep{Verma:book:ET}
\be
S^{\omega \omega}({\bf k|p|q}) = \Im \left[   \{ {\bf k} \cdot  {\bf u}({\bf q}, t) \}   \{ \omega ({\bf p}, t)  \omega^*({\bf k}, t) \} \}    \right].
\label{eq:Sww}
\ee
A simplification of Eq.~(\ref{eq:Sww}) yields \citep{Verma:book:ET} 
\bea
S^{\omega \omega}({\bf k|p|q})   =  \frac{kp}{\cos \gamma} S^{u_1 u_1} ({\bf k|p|q}). 
\label{eq:Sww_Suu}
\eea
In terms of the transformed variables $u,v,w$ of Eq.~(\ref{eq:uvw_transform}), we obtain
\bea
S^{\omega \omega}({\bf 1|v|w})   =  \frac{v}{z} S^{u_1 u_1} ({\bf 1|v|w}) .
\label{eq:Sww_Suu_vw}
\eea
\begin{figure}
	\begin{center}
		\includegraphics[scale = 1]{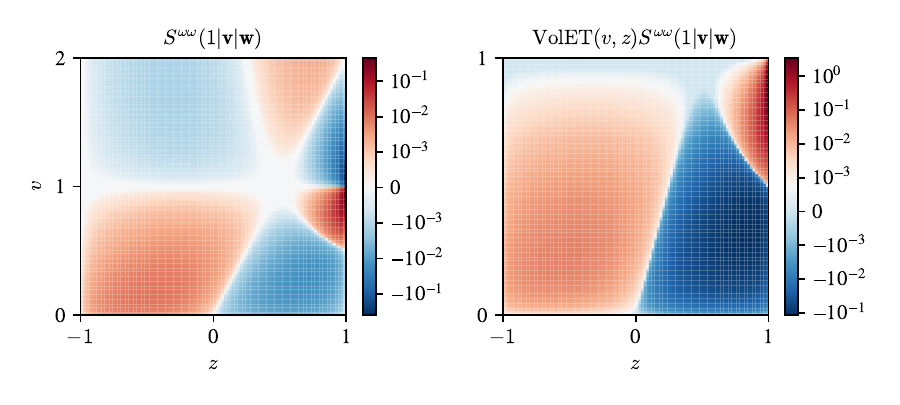}
	\end{center}
	\vspace*{0pt}
	\caption{For enstrophy transfers in 2D HDT: (a) Density plot of $\la S^{\omega \omega}({\bf k'|p|q}) \ra$. (b)  Density plot of $[v \log(1/v)/\sqrt{1-z^2} ] \la S^{\omega \omega}({\bf k'|p|q}) \ra$.}
	\label{fig:Skpq_ww}
\end{figure}

We have already estimated $S^{u_1 u_1} ({\bf 1|v|w}) $ using  field theory (see Section~\ref{sec:ETu1}). We will use these results to estimate $S^{\omega \omega}({\bf 1|v|w})$.   In Fig.~\ref{fig:Skpq_ww}, we illustrate the density plots of $S^{\omega \omega}({\bf 1|v|w})$ and $\mathrm{VolET} S^{\omega \omega}({\bf 1|v|w})$.  It is interesting to contrast
 Fig.~\ref{fig:Skpq_CH1}  and Fig.~\ref{fig:Skpq_ww}.

Triads of Type A and B of Fig.~\ref{fig:triads} have small $S^{\omega \omega}({\bf 1|v|w})$ due to its proportionality with $S^{u_1 u_1} ({\bf 1|v|w})$.   However, triads of Type C, D, and E exhibit  large magnitudes for $S^{\omega \omega}({\bf k|p|q})$, and their asymptotic hehaviour are as follows.  For triads with $v \approx 1$ and $ z \approx 1$ (Types C and D), using Eq. (\ref{eq:Skpq2D_asymptot_typeCD}), we derive that 
\be
\la S^{\omega \omega}({\bf 1|v|w}) \ra   \approx \la S^{u_1 u_1}({\bf 1|v|w}) \ra   \approx  \frac{4}{3\pi^2}  w^{-14/3} (1-v)(1-z).
\ee

For the triads with $v \approx 0$ (Type E), $\la S^{\omega \omega} ({\bf 1|v|w}) \ra $ and 
$\la S^{u_1 u_1} ({\bf 1|v|w}) \ra$  have  very different properties.  Following Eq.~(\ref{eq:2D_v_ll_1_stage0}), we derive that
\bea
\la S^{\omega \omega}({\bf 1|v|w}) \ra   & = & \frac{v}{z} \la S^{u_1 u_1}({\bf 1|v|w}) \ra   \nonumber \\
& \approx &  \frac{1}{ \pi^2}  
\left[  -\frac{2}{3}  v^{1/3} z + v^{4/3} \left(\frac{16}{3} z^2 -1 \right) \right] (1-z^2).
\label{eq:Sww_v_0}
\eea
To a leading-order,  Eq.~(\ref{eq:Sww_v_0}) yields
\be
\la S^{\omega \omega}({\bf 1|v|w}) \ra \approx - \frac{2}{3\pi^2}  v^{1/3} z
\ee 
that satisfies the following property:
\be
\la S^{\omega \omega}({\bf 1|v|w}) \ra (v,z) = - 
\la S^{\omega \omega}({\bf 1|v|w}) \ra (v,-z) .
\ee
Hence, 
\be
\la S^{\omega \omega}({\bf 1|v|w}) \ra 
= \begin{cases}
	> 0 & \mathrm{for}~z < 0 \\
	< 0 & \mathrm{for}~  z > 0
\end{cases}
\ee	
which is unlike $\la S^{u_1 u_1}({\bf 1|v|w}) \ra$ where the contributions from all of small $z$'s add up.  Figures~\ref{fig:Skpq_CH1} and \ref{fig:Skpq_ww} clearly illustrate the above contrast the small $z$.   These contrasting behaviour is responsible for the inverse energy cascade and the forward, but vanishing, enstrophy flux in the $k^{-5/3}$ regime of 2D HDT~\citep{Boffetta:ARFM2012}.

Note, however, that $\int_{-1}^1 z dz =0$. Hence, the integral of the leading-order  term of   Eq.~(\ref{eq:Sww_v_0}) vanishes.  However,  the integral of $\la S^{\omega \omega}({\bf 1|v|w}) \ra (v,z) $ is nonzero due to the higher-order terms of Eq.~(\ref{eq:Sww_v_0}).

\subsection{Enstrophy Flux}
The enstrophy flux for a wavenumber sphere of radius $R$ is defined as~\citep{Lesieur:book:Turbulence,Verma:book:ET}
\be
\Pi_\omega(R) =\int_{R}^\infty \frac{d{\bf k'}}{(2\pi)^2}  \int_0^{R} \frac{d{\bf p}}{(2\pi)^2} \la S^{\omega \omega}({\bf k'|p|q}) \ra .
\ee
Using the relation of Eq.~(\ref{eq:Sww_Suu}) and Eqs.~(\ref{eq:Pi_2D_integral0}, \ref{eq:Pi_2D_integral}), we derive a field-theoretic estimate of the enstrophy flux as
\bea
\frac{ \la \Pi_\omega(k)\ra}{\epsilon_u }   =  k^2  \frac{ K_\mathrm{Ko2D}^{3/2}}{ \nu_{1*}}  4\pi
\int_{0.256}^1 dv   [\log(1/v)] v \int_{-1}^1  \frac{dz}{\sqrt{1-z^2}}  \frac{v}{z} \frac{\mathrm{numr}_2}{\pi^2(1+v^{2/3} + w^{2/3})) }
\eea
with $\mathrm{Ko2D} = 6.013$ and $ \nu_{1*} = 0.1441$. We choose the lower limit for the $dv$ integral as 0.22, as in the energy flux computation that yields $\mathrm{Ko2D} = 6.395$. We employ the Gauss-Jacobi quadrature for the $dz$ integral, and a Romberg scheme for the $dv$ integral.  Our integral computation yields
\be
\frac{ \la \Pi_\omega(k)\ra}{\epsilon_u }  \approx  k^2
\frac{ K_\mathrm{Ko2D}^{3/2}}{ \nu_{1*}}   0.02242 
\approx  3.534 k^2.
\label{eq:Pomega_k2}
\ee
\begin{figure}
	\begin{center}
		\includegraphics[scale = 0.5]{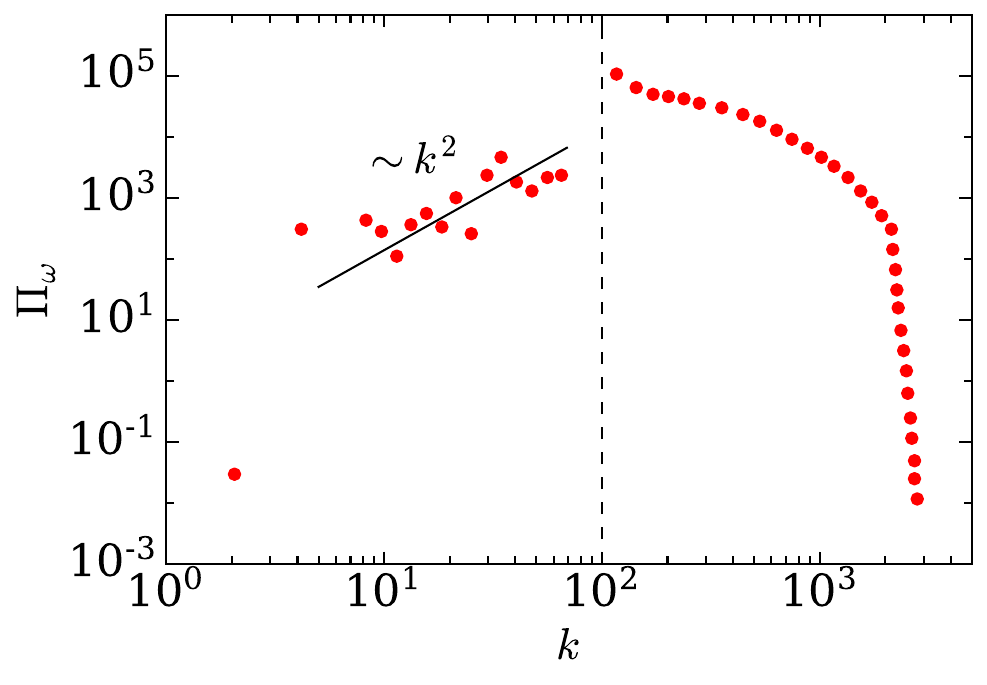}
	\end{center}
	\vspace*{0pt}
	\caption{For 2D HDT with $k_f = 100$,  plot of enstrophy flux $\Pi_\omega(k)$ vs.~$k$. We observe that $\Pi_\omega(k) \sim k^2$ for $k<k_f$, consistent with the predictions of Eq.~(\ref{eq:Pomega_k2}). Adopted from Fig.~3(f) of \cite{Gupta:PRE2019}. Reprinted with permission from APS. }
	\label{fig:Pi_omega}
\end{figure}

Interestingly, the enstrophy flux is positive, but it decreases as $k \rightarrow 0$. The opposite signs for the enstrophy and energy fluxes are consistent with  \cite{Fjortoft:Tellus1953}'s argument  that in 2D HDT, the energy and enstrophy cascade along different directions. \cite{Gupta:PRE2019} simulated 2D HDT and computed the enstrophy flux for a forced case.  Figure~\ref{fig:Pi_omega} illustrates $\Pi_\omega(k)$ for one of their runs. The figure illustrates that $\Pi_\omega(k) \sim k^2$, consistent with the predictions of Eq.~(\ref{eq:Pomega_k2}).  For 3D HDT, \cite{Verma:book:ET} employed scaling arguments and showed that $\la \Pi_\omega(k)\ra \sim k^2$. \cite{Sadhukhan:PRF2019} verified the above results using numerical simulations.

We expect certain complications arising due to wavenumber-dependent ($k^2$) enstrophy flux, $\Pi_\omega(k)$~\citep{Gupta:PRE2019,Verma:JPA2022}.  We will address this issue in a future communication.

%\bibliographystyle{jfm}
%\bibliography{/Users/mkv/Dropbox/docs-pub/bib/journal,/Users/mkv/Dropbox/docs-pub/bib/book,/Users/mkv/Dropbox/docs-pub/bib/book_chapter,/Users/mkv/Dropbox/docs-pub/bib/conf,/Users/mkv/Dropbox/docs-pub/bib/thesis,/Users/mkv/Dropbox/docs-pub/bib/report}

\end{document}